\documentclass[twocolumn]{aastex62}
\usepackage{graphicx}
\usepackage{longtable}
\usepackage{amsmath}
\usepackage{ulem}
\newcommand*{\courier}{\fontfamily{pcr}\selectfont}
\DeclareTextFontCommand{\textmyfont}{\courier}

\begin{document}

\title{The ELM Survey South. I. An Effective Search for Extremely Low Mass White Dwarfs}

\author{Alekzander Kosakowski}\affiliation{Homer L. Dodge Department of Physics and Astronomy, University of Oklahoma, 440 W. Brooks St., Norman, OK, 73019 USA}
\author{Mukremin Kilic}\affiliation{Homer L. Dodge Department of Physics and Astronomy, University of Oklahoma, 440 W. Brooks St., Norman, OK, 73019 USA}
\author{Warren R. Brown}\affiliation{Smithsonian Astrophysical Observatory, 60 Garden St, Cambridge, MA 02138 USA}
\author{Alexandros Gianninas}\affiliation{Department of Physics and Astronomy, Amherst College, 25 East Drive, Amherst, MA 01002, USA}

\begin{abstract}

We begin the search for extremely-low mass ($M\leq0.3M_{\odot}$, ELM) white dwarfs (WDs) in the southern sky based on photometry from
the VST ATLAS and SkyMapper surveys. We use a similar color-selection method as the Hypervelocity star survey.  We switched to an astrometric selection once Gaia Data Release 2 became available. We use the previously known sample of ELM white dwarfs to demonstrate that these objects occupy a unique parameter space in parallax and magnitude. We use the SOAR 4.1m telescope to test the Gaia-based selection,
and identify more than two dozen low-mass white dwarfs, including 6 new ELM white dwarf binaries with periods as short as 2 h.
The better efficiency of the Gaia-based selection enables us to extend the ELM Survey footprint to the southern sky. 
We confirm one of our candidates, J0500$-$0930, to become the brightest ($G=12.6$ mag) and closest ($d=72$ pc) ELM white dwarf binary currently known.
Remarkably, the Transiting Exoplanet Survey Satellite (TESS) full-frame imaging data on this system reveals
low-level ($<0.1$\%) but significant variability at the orbital period of this system ($P=9.5$ h), likely from the relativistic beaming effect.
TESS data on another system, J0642$-$5605, reveals ellipsoidal variations due to a tidally distorted ELM WD.
These demonstrate the power of TESS full-frame images in confirming the orbital periods of relatively bright compact object binaries.

\end{abstract}

\section{Introduction}

The single-star evolution of a solar-metallicity main sequence star with mass below about 8 M$_\odot$ typically results in the formation of a CO-core white dwarf with mass of around 0.6-0.8 M$_\odot$ or an ONe-core white dwarf with mass of around 1.0 M$_\odot$ \citep{Woosley2015,Lauffer2018}. The formation of low mass He-core white dwarfs ($M<0.5$ M$_\odot$) requires that the progenitor loses a significant amount of mass while on the red giant branch. This mass loss can occur in metal-rich single-stars \citep{kilic2007} or in close binary systems, in which the companion strips the low-mass white dwarf progenitor of its outer envelope before it begins Helium burning.

Extremely Low Mass white dwarfs (ELM WDs) are a relatively rare population of $M\leq0.3 {\rm M}_\odot$ He-core white dwarfs that form after severe mass loss. Because the main sequence lifetime of an ELM WD progenitor through single-star evolution is longer than a Hubble time, these ELM WD systems must form through binary interaction, typically following one of two dominant evolutionary channels: Roche lobe overflow or common-envelope evolution \citep{li2019}. While almost all of the known ELM WD systems are found in compact binaries, \citet{justham2009} predict a population of single ELM WDs that are the surviving cores of giant stars whose envelope was stripped by a companion during a supernova explosion.

In support of binary evolution models, virtually all known ELM WDs are found in binary systems, with about half of the known systems expected to merge within a Hubble time due to the emission of gravitational waves \citep{kilic10,brown10,Brown2020}. Compact double-degenerate merging systems are the dominant sources of the gravitational wave foreground at mHz frequencies \citep{nelemans01,nissanke12,korol17,lamberts19}. Identification of additional merging systems allows for better characterization of the gravitational wave foreground for the upcoming Laser Interferometer Space Antenna (LISA) mission. At the time of writing, three of the strongest seven LISA calibration sources are compact double-degenerate binaries, all of which contain an ELM WD \citep{Brown2011,Kilic2014,Burdge2019}.

\begin{figure}[htbp]
\centering
\vspace{-50pt}
\includegraphics[scale=0.45]{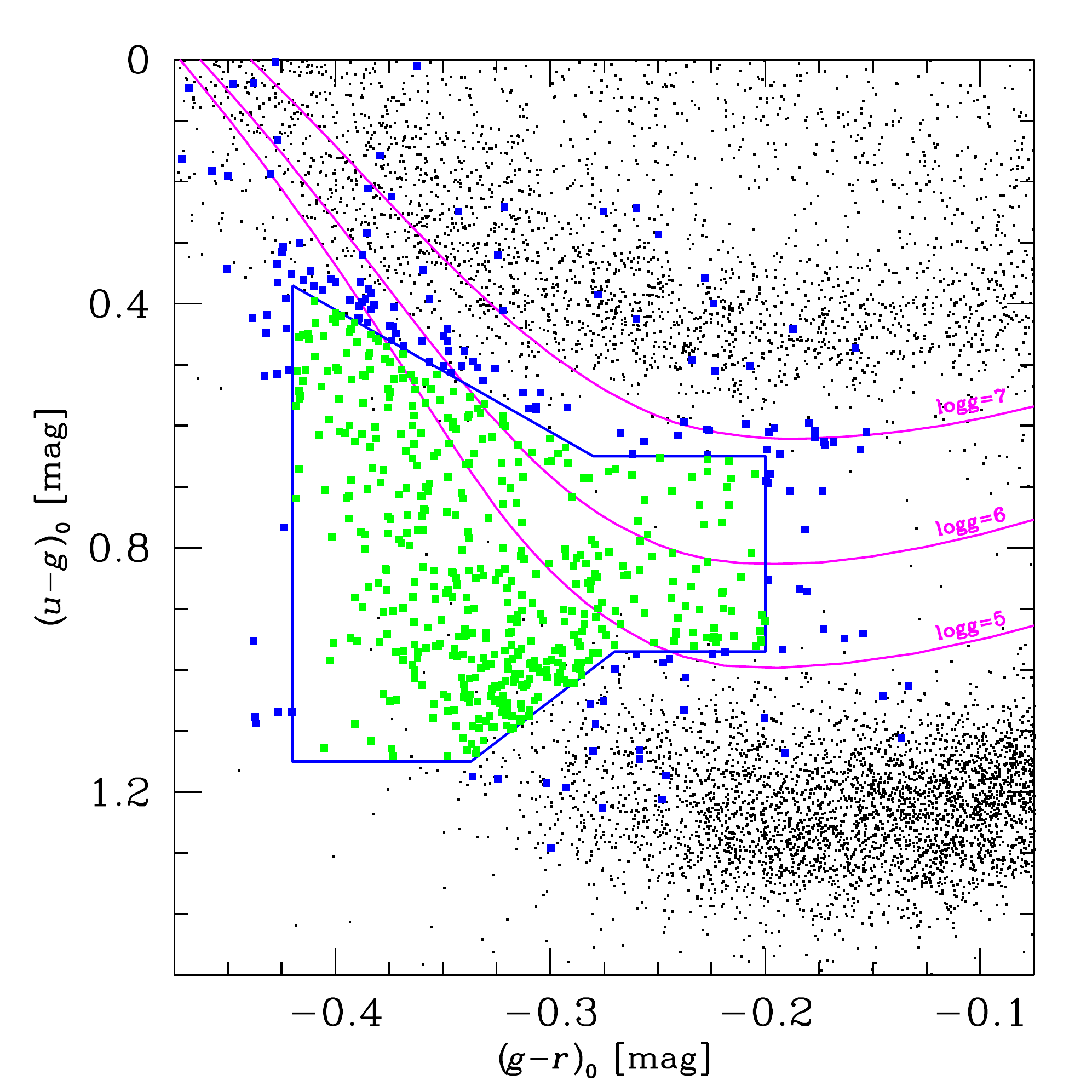}
\vspace{-20pt}
\caption{Target selection region for the VST ATLAS data set as described in the text.
The colored dots mark every ATLAS object with $15<g_0<20$ mag and with a follow-up
spectrum (green points satisfy the target selection region shown in blue bounds, and
blue points are outside of our final target list), black dots are all other
objects in ATLAS, restricted to $17.5<g<19.5$ mag for sake of clarity. We overplot
the DA white dwarf cooling tracks for $\log{g}$=5,6,7 as magenta lines.
Excluding objects visually identified as ``bad'' (close doubles, objects
in globular clusters, etc.), our spectroscopic follow-up is 89\% complete
in the range $16<g_0<20$ mag.}
\label{atlas_selection_plot}
\end{figure}

The fate of ELM WD systems is strongly dependent on the mass ratio of the stars in the system. The system's mass ratio determines whether eventual mass transfer is stable or unstable \citep{marsh04,kremer17}, which then determines the system's merger timescale and merger outcomes. Potential outcomes for these merging systems include single massive white dwarfs, supernovae, Helium-rich stars such as R CorBor stars, and AM CVn systems. While it is generally thought that stable mass transfer results in an AM CVn, \cite{Shen2015} have shown that, through dynamical friction caused by nova outbursts, all interacting double-degenerate white dwarf systems may merge \citep[see also][]{brown16}. To fully understand the formation channels of these various merger outcomes, a more complete sample of merging progenitor systems is needed. Because ELM WD systems are signposts for compact binary systems, increasing the ELM WD sample directly improves the sample of merging systems.

Previous surveys targeting ELM WDs have taken advantage of the abundance of photometric measurements of the northern sky to select candidate systems for follow-up observations. At the conclusion of the ELM Survey, \cite{Brown2020} had identified 98 double-degenerate white dwarf binary systems through careful photometric cuts in SDSS photometry, which account for over half of the known double-degenerate systems in the Galaxy.
With almost all of the currently known ELM systems located in the northern sky, we begin the search for ELM systems in the southern sky with two different target selection methods based on ATLAS, SkyMapper, and Gaia photometry.

The layout of this paper is as follows. In section 2, we begin by discussing our ATLAS+SkyMapper color target selection method and observations. We discuss results and briefly comment on the detection efficiency. In section 3, we discuss our Gaia parallax target selection method and discuss the results and efficiency. Finally, we summarize our conclusions in section 4.

\section{A Survey Based on ATLAS and SkyMapper Colors}

The ELM Survey has been successful at identifying a large number of double white dwarfs based on the Sloan Digital Sky Survey
(SDSS) photometry. The $u-g$ and $g-r$ colors are excellent indicators of surface gravity and temperature, respectively. 
With the availability of the $u-$band data from the VST ATLAS and SkyMapper surveys in the southern sky, we based
our target selection on color cuts to the VST ATLAS Data Release 2 and Data Release 3 \citep{Shanks2015} and SkyMapper
Data Release 1 \citep{wolf18}.

\subsection{ATLAS Color Selection}

VST ATLAS is a southern sky survey designed to image 4,500 deg$^2$ of the southern sky at high galactic latitudes in the SDSS $ugriz$ filter set with similar limiting magnitude to SDSS ($r\sim22$). With the release of DR3 in March 2017, each filter has a total southern sky coverage of $\approx3,000-3,700$ deg$^2$.

We constructed our color cuts based on the results of the previous ELM WD \citep{Brown2016} and Hypervelocity Star \citep{Brown2014} surveys. We defined our color cuts to include the region of color-space including late-B type hypervelocity star candidates, which coincidentally overlaps with the low-mass white dwarf evolutionary tracks. Figure \ref{atlas_selection_plot} shows our color selection region.

\begin{figure}[htbp]
\centering
\vspace{-50pt}
\includegraphics[scale=0.45]{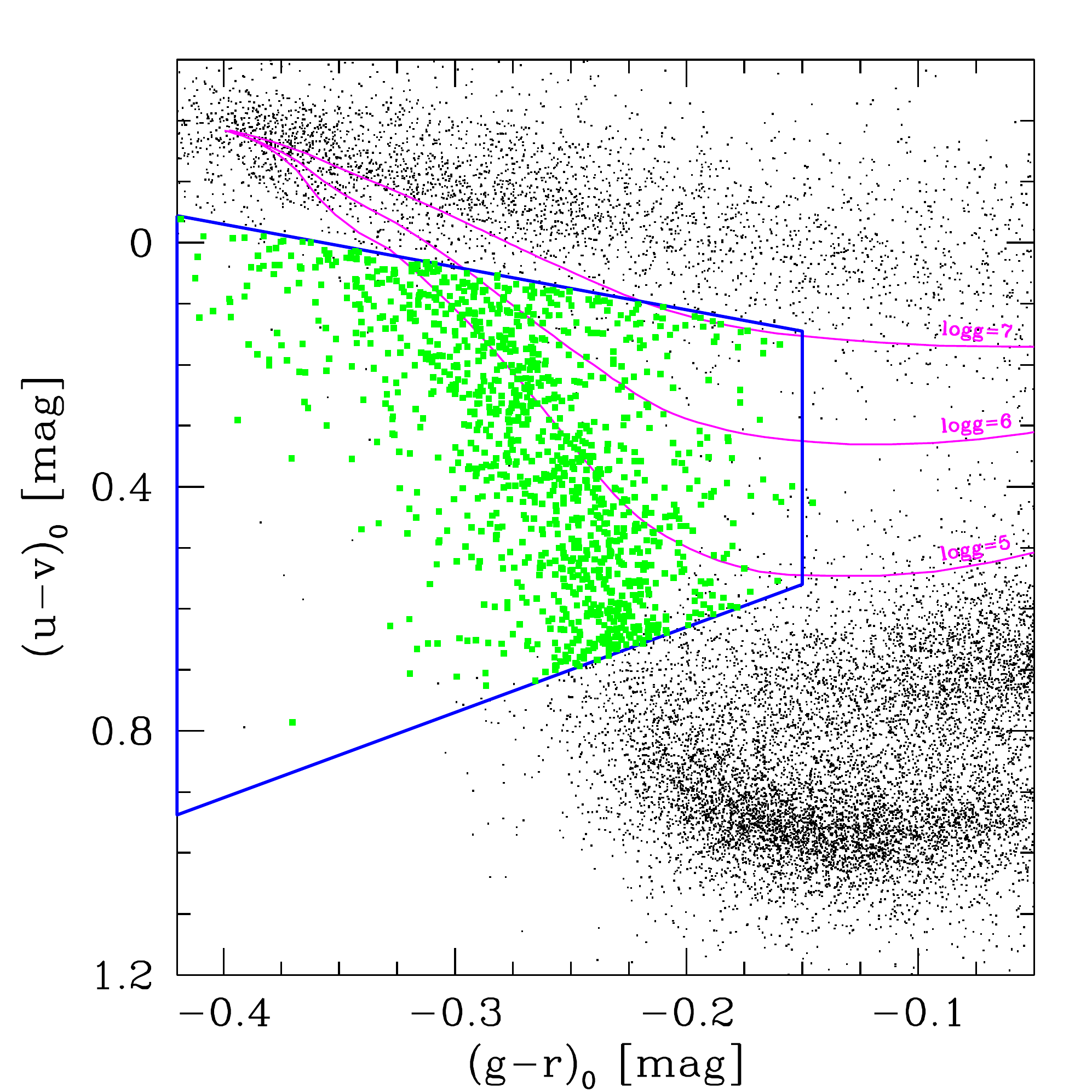}
\vspace{-20pt}
\caption{Target selection region for the SkyMapper DR1 data set as described in the text.
Blue lines mark our target selection region.
Green points represent all of our Skymapper candidates. We overplot
the DA white dwarf cooling tracks for $\log{g}$=5,6,7 as magenta lines.}
\label{skymapper_selection_plot}
\end{figure}
 
\begin{deluxetable*}{llccccc}
\tablecolumns{7}
\tablewidth{0pt}
\tablecaption{Observing setup summary for our ATLAS + SkyMapper observations.}
\tablehead{
\colhead{Telescope} & \colhead{Instrument} & \colhead{Grating} & \colhead{Slit} & \colhead{Resolution} & \colhead{Spectral Coverage} & \colhead{\# Targets Observed} \\
\colhead{} & \colhead{} & \colhead{(lines mm$^{-1}$)} & \colhead{} & \colhead{(\AA)} & \colhead{(\AA)} & \colhead{}
}
\startdata
{SOAR 4.1m} & {Goodman} & {930} & {0.95\arcsec} & {2.4} & {$3550-5250$} & {48} \\
{} & {} & {} & {1.01\arcsec} & {2.6} & {$3550-5250$} & {487} \\
{Walter Baade 6.5m} & {MagE} & {175} & {0.70\arcsec} & {1.0} & {$3600-7000$} & {134} \\
{Tillinghast 1.5m} & {FAST} & {600} & {1.50\arcsec} & {1.7} & {$3600-5400$} & {10} \\
{} & {} & {} & {2.00\arcsec} & {2.3} & {$3600-5400$} & {2} \\
{MMT 6.5m} & {Blue Channel} & {832} & {1.00\arcsec} & {$1.0$} & {$3500-4500$} & {21} \\
{} & {} & {} & {1.25\arcsec} & {$1.2$} & {$3500-4500$} & {10} \\
\enddata
\vspace{1cm}
\end{deluxetable*}

To construct our VST ATLAS DR2+DR3 sample, we first de-reddened and converted the native ATLAS colors into SDSS ($u_0$, $g_0$, $r_0$, and $i_0$) using reddening values of \cite{Schlegel1998} and color conversion equations of \cite{Shanks2015}. We exclude targets located along the line of sight to the Galactic bulge and restricted target $g_0$ magnitude to $15\leq g_0 <20$. We remove quasars from the list
by imposing a cut on $r-i$, and limit our sample to objects with $11,000 {\rm\ K} \lesssim T_{\rm eff} \lesssim 22,000 {\rm\ K}$
by imposing a $g-r$ color cut. While our temperature limits are chosen to avoid contamination from sdA
and sdB stars, which are generally found outside of this temperature range,
such a temperature cut introduces a selection bias against ELM WD systems
that form through stable Roche lobe overflow \citep{li2019}. Our exact photometric cuts are defined by
\begin{flalign*}
15 &\leq g_0 < 20 \\
-0.42 &< (g-r)_0 < -0.2 \\
(r-i)_0 &< -0.05 \\
(u-g)_0 &< 1.15 \\
(u-g)_0 &< -2.67(g-r)_0 + 0.25\ ||\ (u-g)_0 < 0.97 \\
(u-g)_0 &> 2.0(g-r)_0 + 1.21\ ||\ (u-g)_0 > 0.65 \\
\end{flalign*}

\newpage
\subsection{SkyMapper Color Selection}

SkyMapper is a southern sky survey designed to image the entire southern sky in the $uvgriz$ filter set. SkyMapper DR1, released June 2017, provides data on over 20,000 deg$^2$ of the southern sky, with approximately 17,200 deg$^2$ covered by all six filters. SkyMapper DR1 is a shallow survey with limiting magnitude around 17.75 for each filter.

From the SkyMapper DR1 dataset, we selected all objects with $E(B-V)<0.1$ and stellarity index $\textmyfont{class\_star}>0.67$, where $\textmyfont{class\_star}=1.0$ represents a star. We then removed targets along the line of sight of the Galactic Bulge and the Large and Small Magellanic Clouds. Finally, we de-reddened and
applied the following color cuts in the native SkyMapper $uvgriz$ system \citep{Bessell2011} to create a clean sample. Figure \ref{skymapper_selection_plot} shows our target selection region.

\begin{flalign*}
&g>10.5 \\
-0.42 < &(g-r)_0 < -0.15\\
0.7(g-r)_0 + 0.25 < &(u-v)_0 < -1.4(g-r)_0 + 0.35 \\
3.5(g-r)_0 + 1.0 < &(u-g)_0 < 0.8 - (g-r)_0 \\
0.91(r-i)_0 - 0.16 < &(g-r)_0 < -0.425(r-i)_0-0.28 \\
\end{flalign*}

\subsection{Observations}

Because the previously known ELM WDs in the main survey \citep{Brown2020} display an average 240 km s$^{-1}$ velocity semi-amplitude, our observation setup is optimized to obtain radial velocity uncertainty
of 10 km s$^{-1}$, which allows for reliable orbital solutions. We initially observed candidates based on color information. We perform atmospheric fits to each target at the end of each night. Targets with atmosphere solutions consistent with ELM WDs are followed up with at least eight radial velocity measurements, including back-to-back exposures and exposures separated by 1 day to search for short and long-period variability. After our initial measurements, we then attempt to sample the fitted RV curve to reduce period aliasing.

Our target selection and observing strategy lead to a bias against
the ELM WDs that form through the stable Roche Lobe overflow channel
 \citep[see figure 10 in][]{li2019}. Some of these are predicted to be found in
longer period systems with lower velocity semi-amplitudes. Our
observing strategy works well for the ELM WDs that we discover, but
we are less likely to find the longer period systems by design. 

A summary of our observing setup for each our ATLAS+SkyMapper target lists is available in Table 1.

We observed 532 unique systems over 14 nights across three observing campaigns on March 2017 (NOAO Program ID: 2017A-0076), August 2017 (NOAO Program ID: 2017B-0173), and March 2018 (NOAO Program ID: 2018A-0233) using the SOAR 4.1-meter telescope located on Cerro Pach\'{o}n, Chile. We used the Goodman high throughput spectrograph \citep{clemens04} with the blue camera and $0.95\arcsec$ or $1.01\arcsec$ slits with 930 lines mm$^{-1}$ grating resulting in spectral resolution of $\approx$2.5\AA\ covering the wavelength range 3550 - 5250\AA, which includes all of the Balmer lines except H$\alpha$. To ensure accurate wavelength calibration, we paired each target exposure with an FeAr or FeAr+CuAr calibration lamp exposure. We obtained multiple exposures of spectrophotometric standard stars each night to facilitate flux calibration. The median seeing for each night ranged from $0.8-1.0\arcsec$.

We observed 134 additional systems using the Walter Baade 6.5-meter telescope with the MagE spectrograph, located at the Las Campanas Observatory on Cerro Manqui, Chile. We used the 0.7$\arcsec$ slit with the 175 lines mm$^{-1}$ grating resulting in spectral resolution of $\approx$1.0\AA\ covering 3,600 - 7,000\AA.

We observed 12 additional systems using the Fred Lawrence Whipple Observatory (FLWO) 1.5-meter Tillinghast telescope with the FAST spectrograph, located on Mt. Hopkins, Arizona. We used the 1.5$\arcsec$ or 2.0$\arcsec$ slits with the 600 lines mm$^{-1}$ grating resulting in spectral resolution of $\approx$1.7\AA\ or $\approx$2.3\AA\ between 3,600\AA\ - 5,400\AA.

We observed 31 additional systems using the MMT 6.5-meter telescope with the Blue Channel Spectrograph, located on Mt. Hopkins, Arizona. We used the 1.0$\arcsec$ or 1.25$\arcsec$ slits with the 832 lines mm$^{-1}$ grating resulting in spectral resolution of 1.0\AA\ or 1.2\AA\ covering the wavelength range 3,500 - 4,500\AA.

\newpage
\subsection{Radial Velocity and Orbital Fits}

We used the IRAF cross-correlation package \citep[RVSAO,][]{Kurtz1998} to calculate radial velocities. For each object, we first
cross-correlated all spectra with a low-mass white dwarf template and then summed them to produce a zero-velocity spectrum unique
to that object. We then measured radial velocities for each exposure against the object-specific zero-velocity template and corrected
for the Solar System barycentric motion. We obtained median radial velocity uncertainty of 10 km s$^{-1}$. To confirm the binary nature
of our candidates, we performed orbital fitting to radial velocity measurements using a Monte Carlo approach based on \cite{Kenyon1986}. 

\subsection{Stellar Atmosphere Fits}

We obtained stellar atmosphere parameters by fitting all of the visible Balmer lines H$\gamma$ to H12 in the summed spectra to
a grid of pure-Hydrogen atmosphere models that cover the range of 4,000 K $\leq T_{\text{eff}}$ $\leq$ 35,000 K and
$4.5\leq\log{g}\leq9.5$ and include Stark broadening profiles of \cite{Tremblay2009}. Extrapolation was performed for targets with temperatures or $\log{g}$ outside of this range. Specifics for our fitting technique
can be found in detail in \cite{Gianninas2011,Gianninas2014}. For the systems in which the Ca II K line is visible, we mask out
the data in the wavelength region surrounding and including the Ca II 3933.66\AA\ line from our fits.
\begin{figure*}[htbp]
\centering
\includegraphics[scale=0.40]{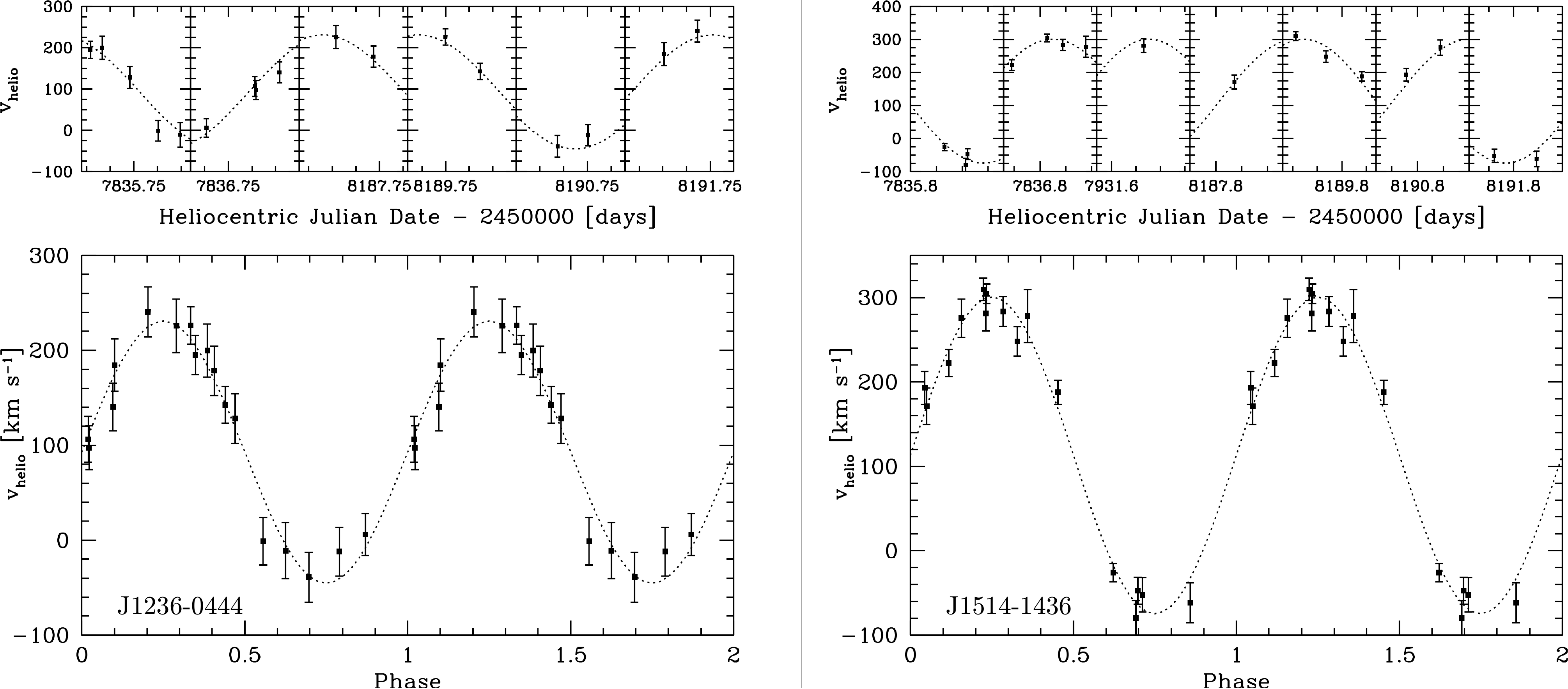}
\caption{Top: Observed radial velocities for J1236-0444 (left) and J1514-1436 (right) with best-fit orbit overplotted as a dotted line.
Bottom: Radial velocity data phase-folded to best-fit period.
A table of radial velocity measurements is available in the Appendix.}
\label{a1236b1514_rv}
\end{figure*}

\begin{figure}[htbp]
\centering
\includegraphics[scale=0.5]{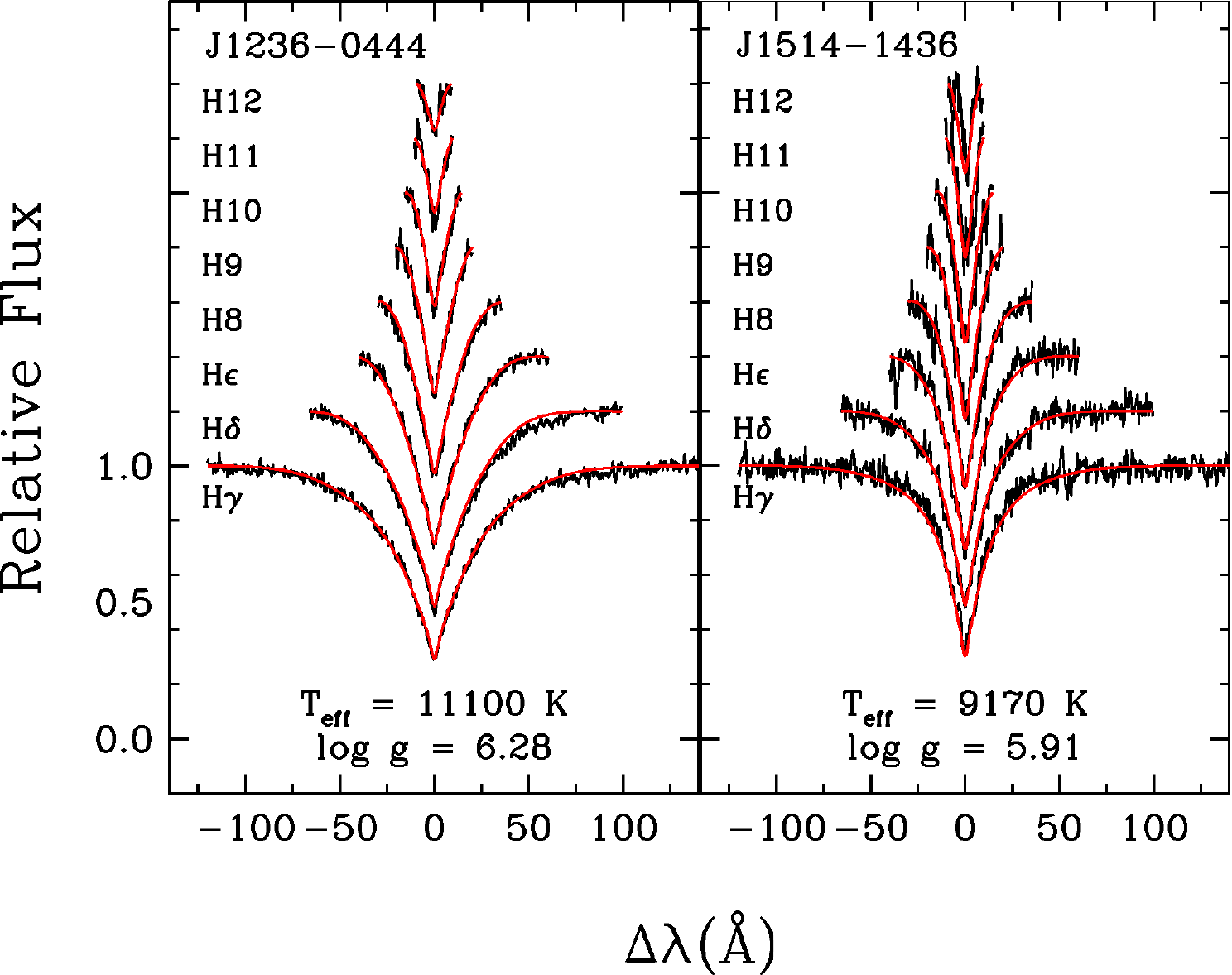}
\caption{Normalized Balmer line profiles for J1236$-$0444 (left) and J1514$-$1436 (right) with best-fitting pure Hydrogen atmosphere model (up to H12) overplotted in red.
Line profiles are shifted vertically for clarity.
The wavelength region surrounding the Ca II 3933.66\AA\ line was masked from the fit to J1514$-$1436.}
\label{a1236b1514_fits}
\end{figure}

\subsection{ELM WDs in ATLAS+SkyMapper}

We fit pure-Hydrogen atmosphere models to all 709 unique targets that show Balmer lines and note that only 33 of these systems are consistent with ELM WD temperature and surface gravity. Of these systems, J0027$-$1516 and J1234$-$0228 are previously published ELM WDs \citep{Brown2020,Kilic2011}. We obtained follow-up spectra and constrained the orbit of three of these systems and confirm that two (J123619.70$-$044437.90 and J151447.26$-$143626.77) are ELM WDs, while the third system (J142555.01$-$050808.60) is likely a metal-poor sdA star. We briefly discuss J1425$-$0508 in the following section. Figures \ref{a1236b1514_rv} and \ref{a1236b1514_fits} show our orbital and model atmosphere fits for J1236$-$0444 and J1514$-$1436.

J1236$-$0444 is an ELM WD with best-fit atmosphere solution of $\log{g}=6.28\pm0.02$ and $T_{\text{eff}}$ = 11,100 $\pm$ 110 K. \cite{Istrate2016} He-core ELM WD evolutionary tracks indicate that J1236$-$0444 is a
0.156$\pm$0.01 M$_\odot$ white dwarf. Orbital fits to the 17 radial velocity measurements give a best-fit period of 0.68758 $\pm$ 0.00327 d
with velocity semi-amplitude of 138.0 $\pm$ 6.6 km s$^{-1}$ (Figure \ref{a1236b1514_rv}, left). Using the binary mass function 
\begin{equation}
\frac{(M_{2}\ \text{sin}\ i)^3}{(M_1+M_{2})^2} = \frac{PK^3}{2 \pi G},
\end{equation}

with primary ELM WD mass $M_1$, orbital period $P$, velocity semi-amplitude $K$, and inclination $i=90^\circ$, we calculate the minimum companion mass $M_{2,{\rm min}}=0.37\pm0.04$ M$_\odot$. 

J1514$-$1436 is an ELM WD with best-fit atmosphere solution of $\log{g}=5.91\pm0.05$ and $T_{\text{eff}}$ = 9,170 $\pm$ 30 K. \cite{Istrate2016} He-core ELM WD evolutionary tracks indicate that J1514$-$1436 is a 0.167$\pm$0.01 M$_\odot$ white dwarf. Orbital fits to the 16 radial velocity measurements give a best-fit period of 0.58914 $\pm$ 0.00244 d with velocity semi-amplitude 187.7 $\pm$ 6.6 km s$^{-1}$ (Figure \ref{a1236b1514_rv}, right). The minimum companion mass for this system is 0.64 $\pm$0.06 M$_\odot$.

The orbit of compact double degenerate systems slowly decays due to the loss of angular moment caused by the emission of gravitational waves \citep{Landau1958}. The merger timescale of these systems can be calculated if the mass of each object and their orbital period is known by using the equation
\begin{equation}
\tau_{{\rm merge}}=\frac{(M_1+M_2)^{1/3}}{M_1M_2}P^{8/3}\times10^{-2}\ {\rm Gyr}
\end{equation}

where $M_1$ and $M_2$ are the ELM WD and companion star masses in solar masses, and $P$ is the period in hours. We use Equation 2 together with the minimum companion mass, $M_{2,{\rm min}}$, to estimate the maximum merger time for these systems. Neither J1236$-$0444 nor J1514$-$1436 will merge within a Hubble time.

\subsection{sdAs in ATLAS+SkyMapper}
\begin{figure}[htbp]
\centering
\includegraphics[scale=0.08]{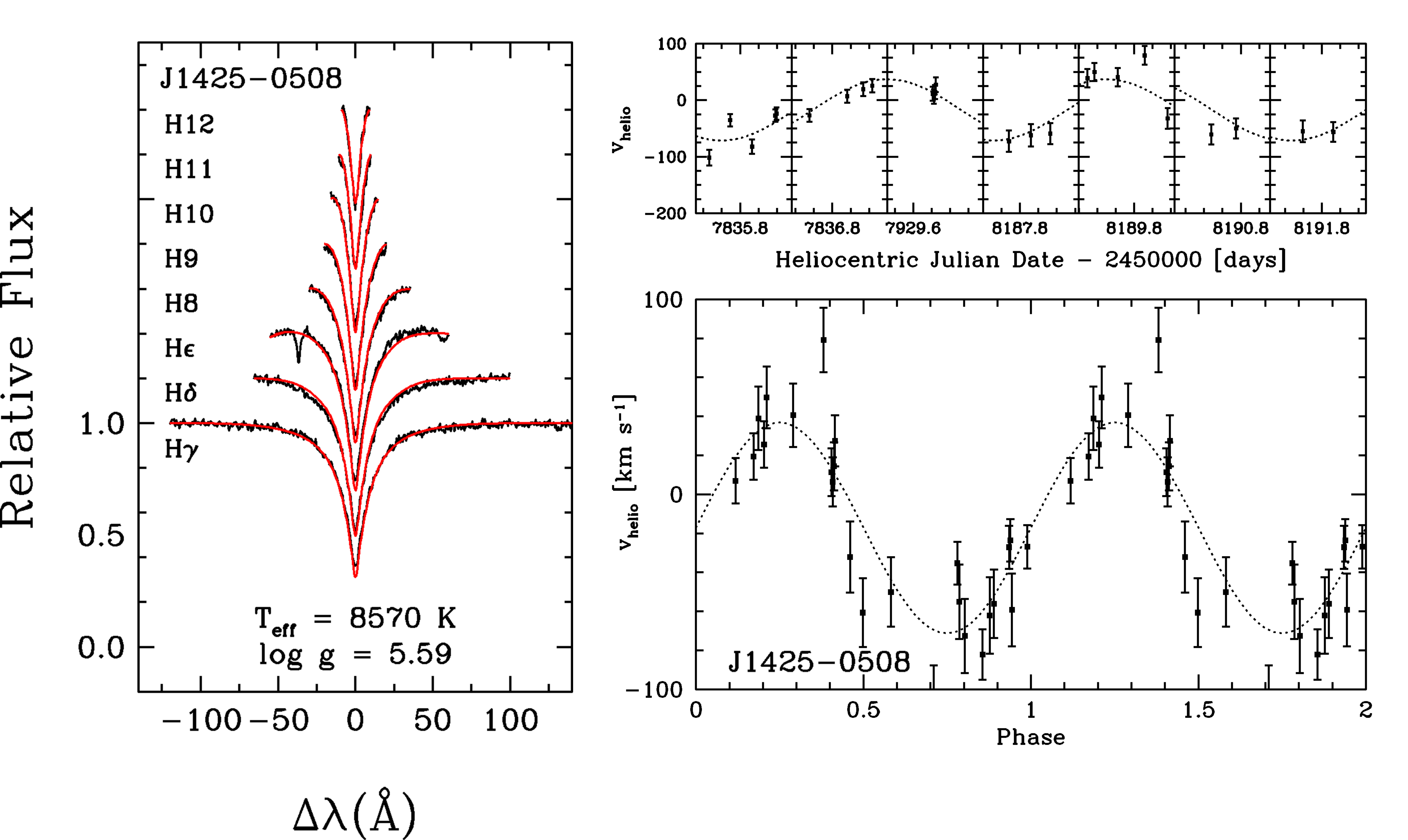}
\caption{Best-fit pure Hydrogen atmosphere model and radial velocity measurements of J1425$-$0508.
A table of radial velocity measurements is available in the Appendix.}
\label{a1425_total}
\end{figure}

In addition to cool ELM WDs, there exists a large population of subdwarf A-type (sdA) stars with $7,000$ K $<T_{{\rm eff}}<20,000$ K (with most below 10,000 K) and $4.5<\log{g}<6.0$ \citep{kepler16,Pelisoli2018a} that are often confused with ELM WDs in low-resolution spectroscopy. \citet{Brown2017} and \citet{Pelisoli2018a,Pelisoli2018b} have shown that the surface gravities derived from pure-hydrogen atmosphere model fits suffer from up to 1 dex error for sdA stars. This is likely due to metal line blanketing that is missing in the pure-hydrogen atmosphere models and the lower signal-to-noise ratio of observed spectra below 3,700 \AA.

We note that while 33 of our objects appear to have atmospheres consistent with ELM WDs, 29 are cool ($T_{\rm eff}<10,000$ K) and share their parameter space with sdA stars. \citet{Yu2019} have shown through binary population synthesis that only 1.5\% of sdA stars in a 10 Gyr old population are ELM WDs, with the remaining 98.5\% being metal-poor main sequence stars \citep[see also][]{Pelisoli2018a,Pelisoli2019}. Therefore, the majority of our 29 candidates with $\log{g}=5-7$ and $T_{\rm eff}=8,000-10,000$ K are likely metal-poor main-sequence stars.

We obtained 25 radial velocity measurements for one of these candidates, J1425$-$0508. Figure \ref{a1425_total} displays our best-fit model atmosphere and orbital fits. J1425$-$0508 is best-explained by a 8,570 K and $\log{g}=5.59$ model based on the assumption of a pure Hydrogen atmosphere. Our radial velocity measurements result in the best-fit period of 0.798 $\pm$ 0.005 d with velocity semi-amplitude $K=54.1\pm3.4$ km s$^{-1}$. As demonstrated by \citet{Brown2017} and \citet{Pelisoli2018a}, the surface gravity for such a cool object is likely over-estimated, and the relatively low semi-amplitude of the velocity variations and the Gaia parallax of 0.25 $\pm$ 0.08 mas favors a low-metallicity main-sequence sdA star, rather than a cool ELM WD.

\begin{figure}[htbp]
\centering
\vspace{-30pt}
\includegraphics[scale=0.57]{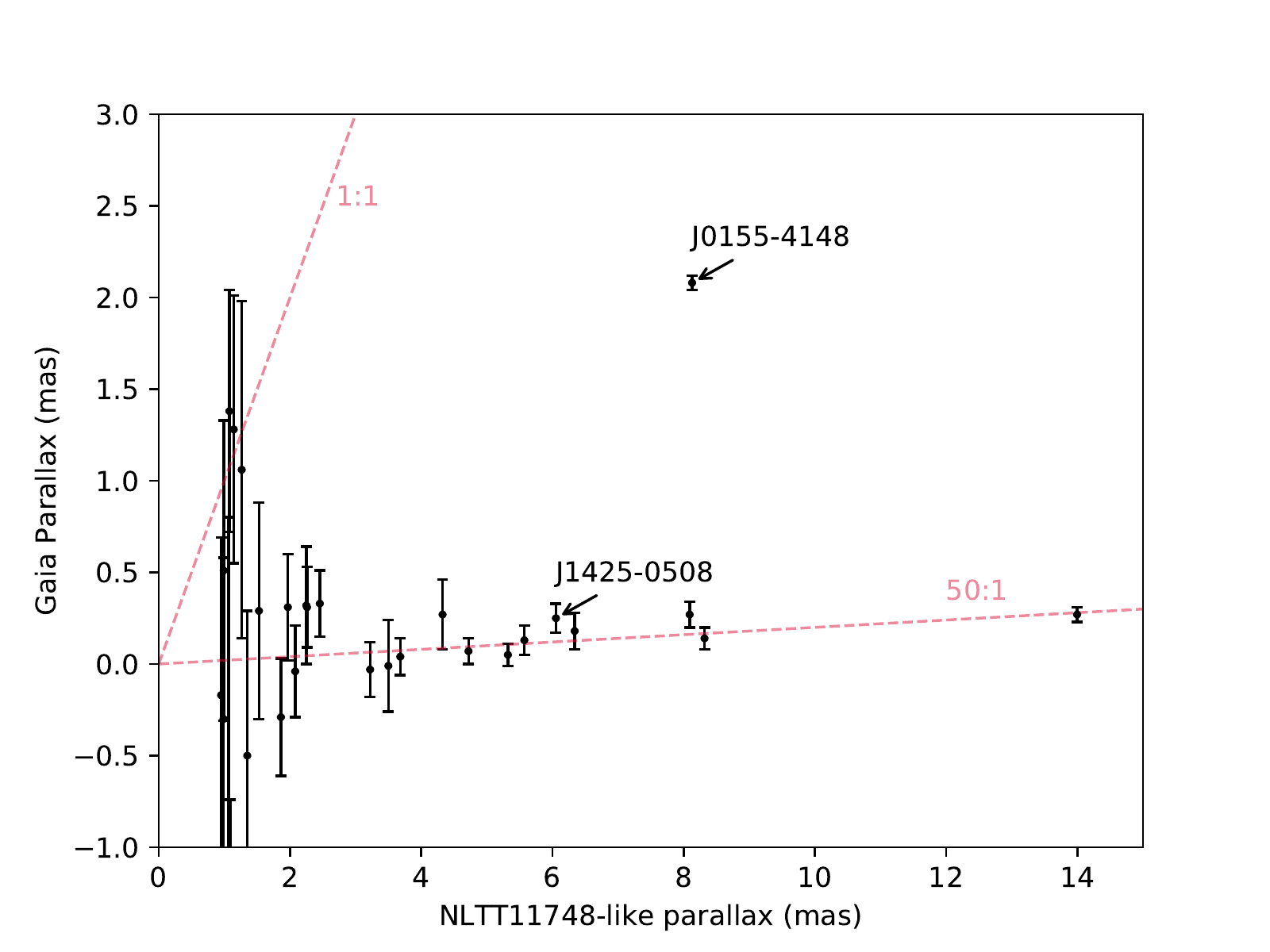}
\caption{Comparison between Gaia parallax and predicted NLTT 11748-like parallax for the 29 sdA stars identified in our survey.
1:1 and 50:1 parallax ratio lines are marked as red dashed lines and labeled.
Candidates consistent with the 50:1 line have a radius estimate of $R\sim2{\rm R}_\odot$ and cannot be white dwarfs.
Candidate J0155$-$4148 lies along the 4:1 line with a radius compatible with an ELM WD.}
\label{nltt11748}
\end{figure}

Given the problems with distinguishing ELM WDs from sdAs, we use the eclipsing system NLTT 11748 \citep{steinfadt2010} as a prototype to estimate the radii of each of our candidates. NLTT 11748 is a well-studied eclipsing ELM WD system with $T_{\rm eff}\approx8,700$ K and $R\approx0.043$ ${\rm R}_\odot$ \citep{Kaplan2014}. We use a similar approach to what is done by \citet{Brown2020} and compare the Gaia parallax for each candidate with its predicted parallax if it were similar in nature to NLTT 11748, obtained by inverting the distance calculated from the candidate's apparent magnitude and the absolute magnitude of NLTT 11748. This comparison provides a radius estimate relative to a known ELM WD.

\begin{figure}[htbp]
\centering
\vspace{-40pt}
\includegraphics[scale=0.70]{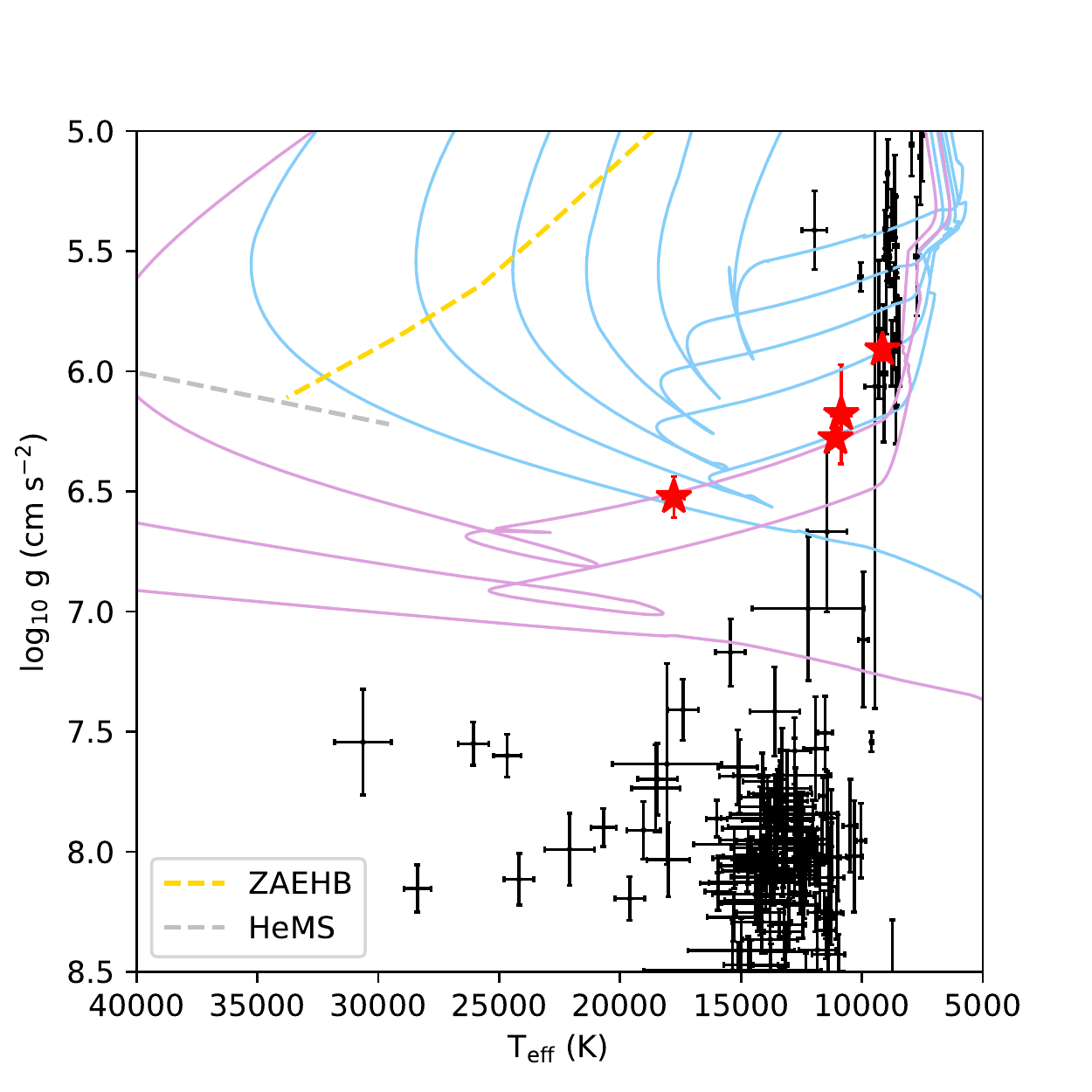}
\caption{$\log{g}$ vs T$_{\rm eff}$ plot of our VST ATLAS DR2+DR3 and SkyMapper DR1 targets with Hydrogen-dominated atmospheres and log$(g)>5.0$.
Red stars show the locations of the confirmed ELM systems identified through our color selection.
Evolutionary tracks for 0.205 M$_\odot$ (light blue) and 0.306 M$_\odot$ (purple) ELM WDs from Istrate et al. (2016) are overplotted.
Hydrogen shell flashes during evolution cause loops seen in the model tracks.
The silver and gold dashed lines show the locations of the Helium main sequence (HeMS) and zero-age extreme-horizontal branch (ZAEHB), respectively.}
\label{atlas_teff_plot}
\end{figure}

Figure \ref{nltt11748} shows the comparison between predicted parallax and Gaia parallax for each of our 29 candidates with the 1:1 and 50:1 lines overplotted. We note that most candidates are consistent with the 50:1 line to within 2$\sigma$, suggesting that they are $\sim$50 times larger than NLTT 11748 with radii $R$$\sim$2 ${\rm R}_\odot$. J0155-4148 is a strong ELM WD candidate; it lies along the 4:1 line with a radius compatible with an ELM WD. We note that there are four additional candidates that are consistent with the 1:1 line, but their Gaia parallax values are uncertain with \textmyfont{parallax\_over\_error<2}. We will present our follow-up observations of J0155$-$4148 in a future publication.

\subsection{Survey Efficiency}

From our ATLAS+SkyMapper color-selection method, we observed 709 unique systems. Of these systems, we confirm only four to contain an ELM WD, two of which were previously known. In addition to these four confirmed ELM WDs, we report 123 DA white dwarfs with $\log{g}>7.0$ (Table 4) and 29 additional candidates with $5.0<\log{g}<7.0$ (Table 5). This low efficiency in our photometric selection may be due to potential color calibration issues in the ATLAS DR3 dataset. In addition, the low efficiency of the SkyMapper selection is likely due to the shallow depth of the SkyMapper DR1, which limits the survey volume for ELM WDs.

Figure \ref{atlas_teff_plot} shows the distribution of temperatures and surface gravities for all targets observed as a part of our ATLAS + SkyMapper color selection with $\log{g}\geq5.0$. We mark the locations of the four observed ELM WD systems with red stars. We overplot the 0.2 M$_\odot$ (light blue) and 0.3 M$_\odot$ (purple) WD evolutionary tracks of \cite{Istrate2016}.

In total, we confirm that only four of our systems (plus the candidate system J0155$-$4148) contain an ELM WD, two of which are new discoveries. Our ATLAS + SkyMapper target selection method has an ELM WD detection efficiency of $\sim$0.6\% and a white dwarf detection efficiency of about 18\%, making the majority of our targets unaligned with our targets of interest.

We note that all four of our confirmed ELM WDs originated from our ATLAS sample. Given the surface density of ELM WDs between $17<g<20$ in the SDSS footprint, we expect to find $\sim$10 ELM WDs in our observed ATLAS sample. However, the spatial distribution of our candidates varied systematically over the ATLAS DR3 footprint, suggesting that photometric calibration in the VST ATLAS DR3 varied across the survey. Similarly for the $15<g<17$ ELM WD sample in the SDSS footprint, we expect to find one ELM WD within our observed SkyMapper sample. While we have not yet confirmed the nature of J0155$-$4148, this system originated from our SkyMapper sample and is likely an ELM WD. Our SkyMapper results are consistent with what is expected given the lower limiting magnitude.

\section{Gaia Parallax Based Selection}

The availability of Gaia DR2 in April 2018 opened a new window into ELM WD target selection.
Gaia photometry and parallax measurements provide a direct measurement of the luminosity of each object, enabling a clear
distinction between low-luminosity WDs and brighter main-sequence stars. ELM WDs are a few times larger in radii compared
to average 0.6 $M_{\odot}$ WDs at the same temperature (color), but they are still significantly smaller than A-type stars. 
Hence, Gaia parallaxes provide a powerful method to create relatively clean samples of ELM WDs \citep[see also][]{Pelisoli2019b},
and also for the first time enable an all-sky survey.

\begin{figure}[htbp]
\centering
\vspace{-20pt}
\includegraphics[scale=0.4]{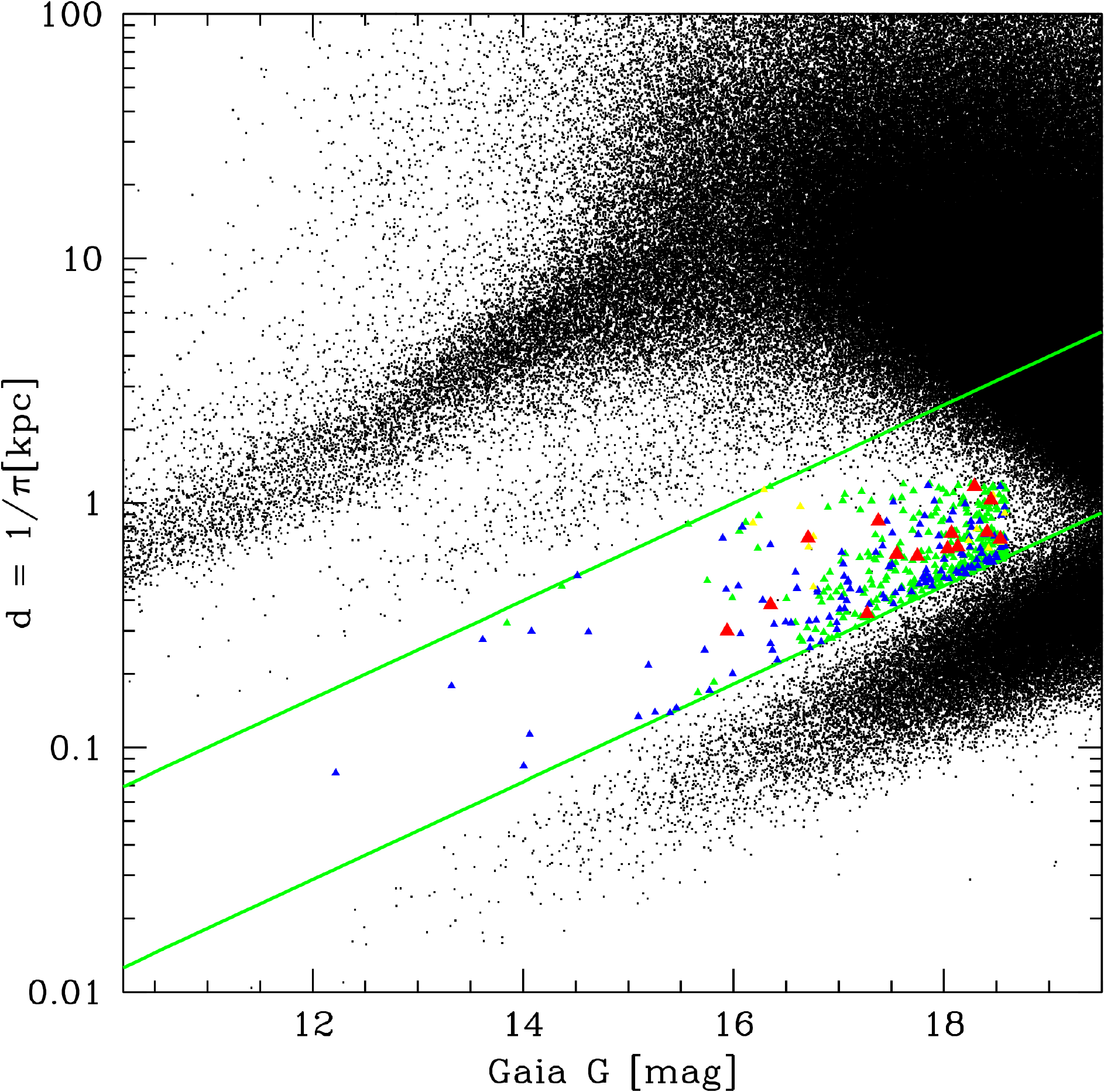}
\caption{Target selection region for Gaia parallax method described in text.
Green lines mark the region for $M_G=6.0-9.7\ {\rm mag}$.
Green triangles are the ELM candidates identified through our Gaia parallax selection.
Red triangles are known ELMs.
Blue triangles are known WDs.
Yellow triangles are other types of previously known objects, like subdwarf B stars and cataclysmic variables.}
\label{gaia_selection_plot}
\end{figure}

Since the ELM Survey has already observed the SDSS footprint, here we focus on the southern sky, but exclude the Galactic plane
($|b| < 20^{\circ}$) due to significant extinction and avoid the Small and Large Magellanic Clouds. We also apply cuts to astrometric noise and color excess based on
recommendations from \cite{Lindegren2018}. Figure \ref{gaia_selection_plot} shows the distances and Gaia magnitudes for
sources with $-0.4 < G_{\rm BP} - G_{\rm RP} < 0.2$. This color range corresponds to $T_{\rm eff} = 8,000 - 25,000$ K, where
Balmer lines are relatively strong. Green lines mark the region for $M_G = 6.0-9.7$ mag objects, and
blue and red triangles mark the previously confirmed normal WDs and ELM WDs in this magnitude range, respectively.
Magenta triangles mark other types of previously known objects, like subdwarf B stars and cataclysmic variables (CVs).

\begin{figure}[htbp]
\centering
\vspace{-20pt}
\includegraphics[scale=0.4]{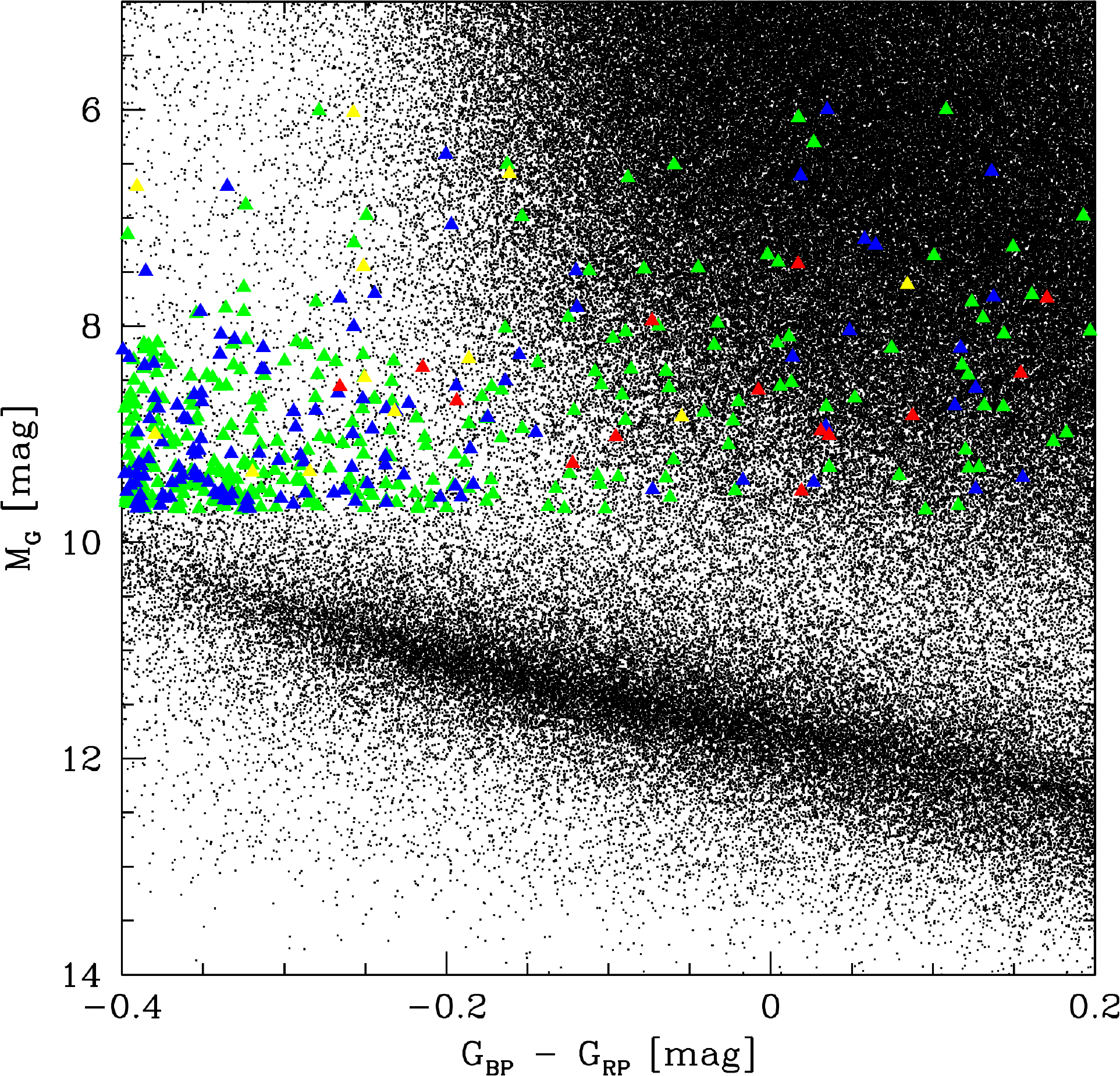}
\caption{Color-magnitude diagram corresponding to our Gaia parallax selection described in the text.
The symbols are the same as in Figure 8. We select objects with Gaia magnitude $M_G = 6.0 - 9.7$ mag. }
\label{gaia_cmd}
\end{figure}

For a more intuitive look at our target selection region, we plot the
same sample on a color-magnitude diagram in Figure \ref{gaia_cmd}. The WD sequence
stretches from $M_G = 10$ mag on the left to about 12 mag on the right.
Our Gaia-selected targets are all over-luminous compared to this sequence
and are dominated by relatively hot WD candidates with bluer colors.
Since we did not impose a cut on parallax errors, the top right portion of
this diagram is dominated by non-WD stars that are scattered
into this region due to large parallax errors.

To minimize contamination from main-sequence stars, we limit our target selection to the region defined by parallax-distance
$(1/\varpi)<1.2$ kpc, and to remove potential contamination from poorly-calibrated colors on fainter targets, we limit the apparent
Gaia G-band magnitude to $G<18.6\ {\rm mag}$. Because normal WDs dominate at larger absolute magnitudes, we impose an
absolute G-band magnitude limit of $M_G<9.7$ to avoid large numbers of normal WDs. Our Gaia target selection resulted in 573 candidates, 180 of which were also identified by \citet{Pelisoli2019b} as ELM WD candidates.

Our Gaia target selection is defined by

\begin{flalign*}
&|b|\geq20 \\
&G < 18.6\ {\rm mag}\\
&6.0<M_G<9.7 \\
&\textmyfont{R.A.} > 100^\circ\ {\rm or}\ (\textmyfont{R.A.}<100^\circ\ \&\ \textmyfont{Dec.} > -60^\circ)\\
&\textmyfont{phot\_bp\_mean\_flux\_over\_error}>10 \\
&\textmyfont{phot\_rp\_mean\_flux\_over\_error}>10 \\
&-0.4\leq(G_{BP}-G_{RP})\leq0.2 \\
&\frac{1}{\textmyfont{$\varpi$}}<1.2 \\
\end{flalign*}

We observed a total of 82 unique systems over four consecutive nights in March 2019 (NOAO Program ID: 2019A-0134). All observations
were taken with the SOAR 4.1-meter telescope using the Goodman Blue Spectrograph with the 1.01$\arcsec$ long-slit resulting
in a spectral resolution of 2.6\AA\ covering the wavelength range of 3550\AA\ - 5250\AA. Median seeing for each night was between
$0.8-1.0$ arcsec. Radial velocities, orbital solutions, and model atmosphere fits were obtained identically to as described in section 2.

\subsection{Results}
\begin{figure}[htbp]
\centering
\includegraphics[scale=0.4]{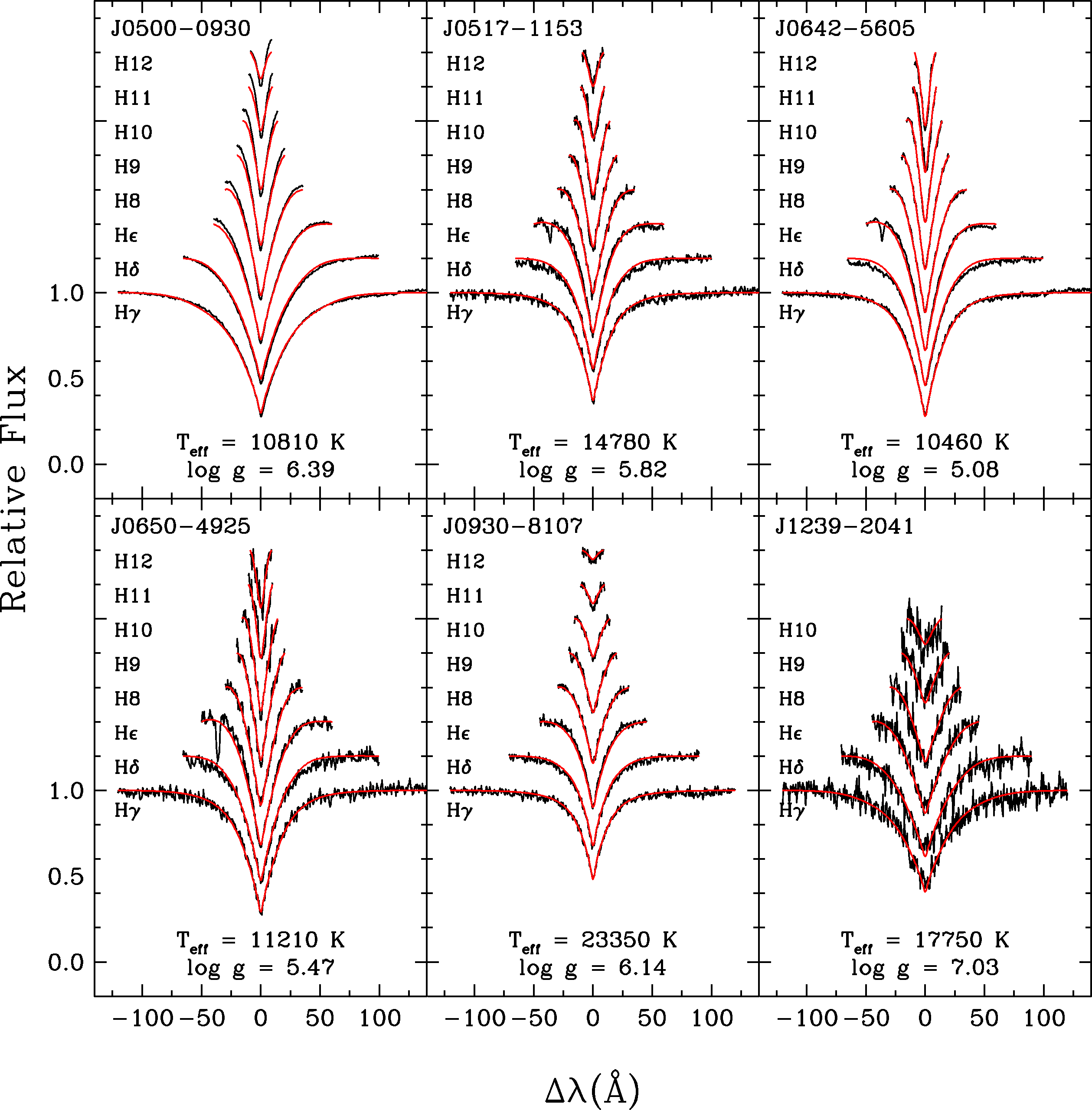}
\caption{Normalized Balmer line profiles for the six new ELM WD systems identified through our Gaia DR2 parallax selection.
Best-fit pure-Hydrogen atmosphere models are overplotted in red with best-fit parameters printed in each subfigure.
The Ca II K line at 3933.66\AA\ in the wing of H$\epsilon$ is masked from fits where it is visible. Line profiles are shifted vertically for clarity.
Due to lower signal-to-noise, we limit our fitting of J1239$-$2041 to include only up to H10.}
\label{elm_fits}
\end{figure}

We fit pure-Hydrogen atmosphere models to all 82 targets and identify six systems consistent with ELM WDs. 
Figure \ref{elm_fits}  shows our model fits to the Balmer line profiles for these six systems. All six are hotter than 10,000 K, have $\log{g}=5-7$, and show significant velocity variability. However, we were only able to constrain the orbital period for
four of these systems so far. Details for each system are discussed below.

\subsection*{{\rm J0500$-$0930}}

J050051.80$-$093056.98 (2MASS J05005185$-$0930549) was originally identified as an ELM WD candidate by \cite{Scholz2018} for its high proper motion. To explain its over-luminous nature, \cite{Scholz2018} suggested that the system contains an ELM WD and estimate atmospheric parameters $\log{g} \approx 6-6.5$ and $T_{\text{eff}}$ = 11,880 $\pm$ 1,100 K.  

\begin{figure}[htbp]
\centering
\vspace{-100pt}
\includegraphics[scale=0.42]{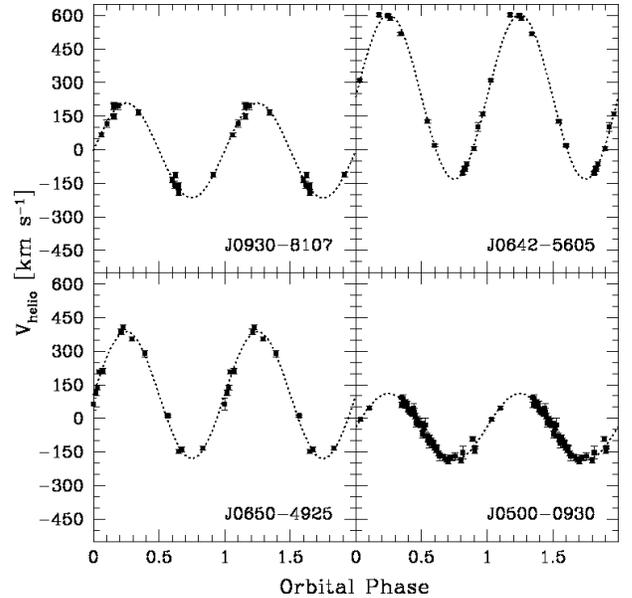}
\caption{Best-fit orbital solutions plotted as a function of phase to the four constrained ELM WD systems based on data from SOAR, FLWO, and MMT.
A table of radial velocity measurements is available in the Appendix.}
\label{elm_rv_fits}
\end{figure}

We obtained $\log{g} = 6.39\pm0.02$ and $T_{\text{eff}}$ = 10,810 $\pm$ 40 K from fitting our SOAR spectra with pure H atmosphere
models, in agreement with the original estimates of \cite{Scholz2018}. We obtained seven radial velocity measurements of J0500$-$0930
with SOAR 4.1-meter telescope using the Goodman spectrograph, 50 with the FLWO 1.5-meter telescope using FAST, and one with
the MMT 6.5-meter telescope with the Blue Channel Spectrograph. Fitting an orbit to this combined dataset of 58 spectra resulted in
a best-fit period of $P=0.39435\pm0.00001$ d with velocity semi-amplitude $K=146.8\pm8.3$ km s$^{-1}$ (Figure \ref{elm_rv_fits}).
 We use the ELM WD evolutionary models of \cite{Istrate2016} to estimate its mass to be 0.163 $\pm$ 0.01 M$_\odot$ and calculated its
 minimum companion mass to be 0.30 $\pm$ 0.04 M$_\odot$, potentially making this a double low-mass WD system. With apparent Gaia
 G-band magnitude of 12.6 and Gaia parallax of 13.97 $\pm$ 0.05 mas, this is currently both the brightest and closest known ELM WD
 system. This system will not merge within a Hubble time.

\begin{figure}[htbp]
\centering
\includegraphics[scale=0.6]{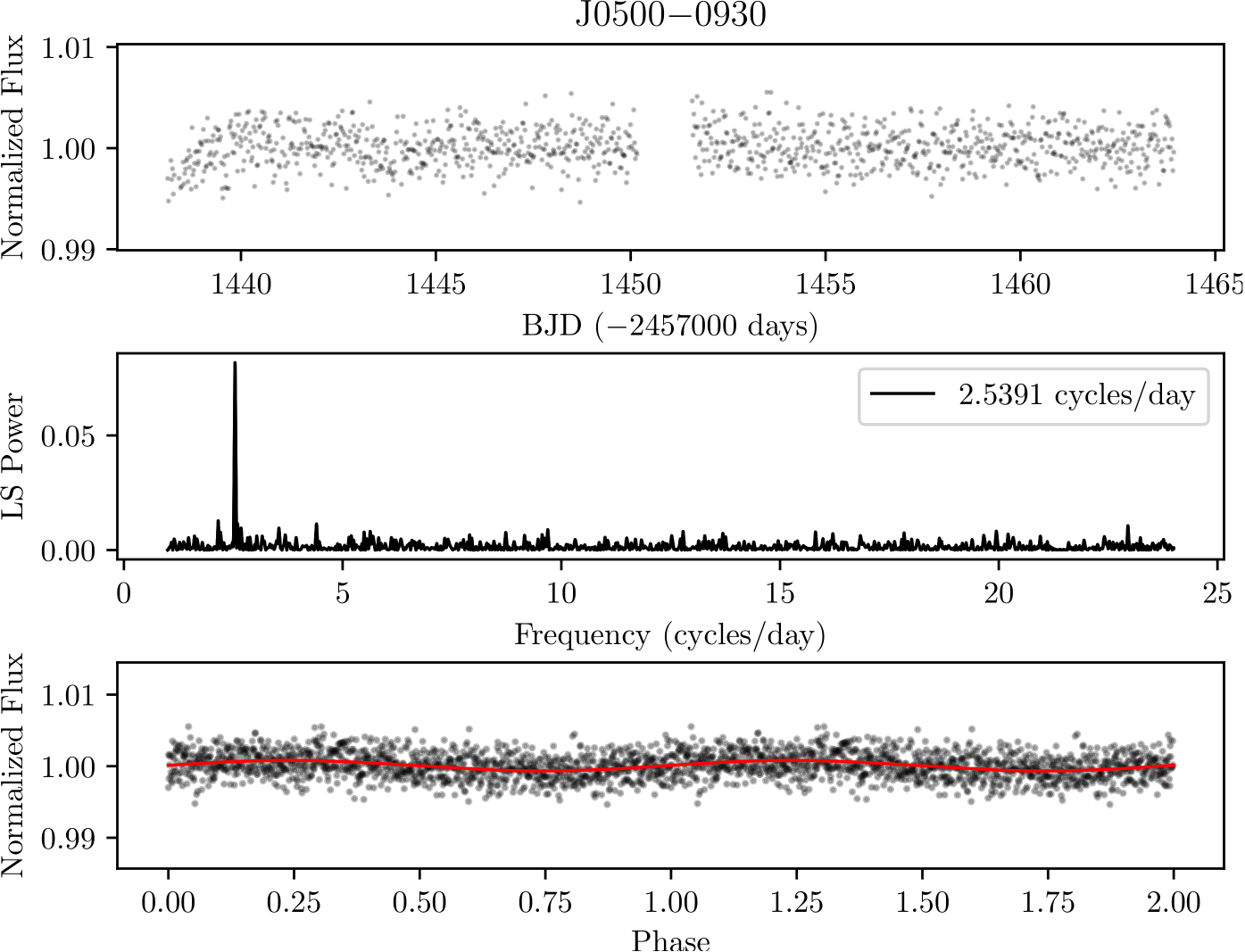}
\caption{TESS Light curve (top), Lomb Scargle periodogram (middle), and phase folded light curve (bottom) of J0500$-$0930. We overplot the best-fit frequency model onto the phase folded light curve for clarity.}
\label{g109_tess}
\end{figure}

J0500$-$0930 was within the field of view of the Transiting Exoplanet Survey Satellite \citep[TESS,][]{ricker15} during Sector 5
observations. TESS provides Full-Frame Images (FFIs) of each sector at 30-minute cadence over a roughly 27 d observing window.
We used the open source Python tool eleanor \citep{feinstein19} to produce a light curve for J0500$-$0930. 
We downloaded a time series of 15 pixels by 15 pixels ``postcards'' containing TESS data for the target and its immediate
surroundings from the Mikulski Archive at the Space Telescope Science Institute (MAST). We then perform background subtraction,
aperture photometry, and correct for instrumental systematic effects. We use the corrected flux measurements with data quality
flags set to 0 to remove data points that are affected by issues like attitude tweaks or cosmic rays \citep{feinstein19}.

We use the Astropy implementation of the Lomb Scargle periodogram to check for variability in the TESS data. Figure \ref{g109_tess} shows the TESS FFI light curve of J0500$-$0930, its Lomb Scargle periodogram, and phase folded light curve at the highest-peaked frequency. Remarkably, there is a small (0.074$^{+0.008}_{-0.007}$\%) but significant peak at a frequency of 2.5391 $\pm$ 0.0025 cycles per day. This frequency is within 1.3$\sigma$ of the orbital frequency measured from our radial velocity data. 
The predicted amplitude of the relativistic beaming effect in J0500$-$0930 is $\sim0.1$\% \citep{Shporer2010}. However,
since the TESS pixels are relatively large (21 arcsec pixel$^{-1}$)
and 90\% of the point spread function is spread over 4 pixel$^2$, dilution
by neighboring sources is common in the TESS data.
There are two relatively red sources with $G_{\rm RP}=16.0$ and $16.9$ mag
within a 2 pixel radius of J0500$-$0930 that likely dilute the variability
signal. Hence, the observed photometric variability is consistent with the relativistic beaming effect,
confirming our orbital period measurement from the radial velocity data.

\subsection*{{\rm J0517$-$1153}}

J051724.97$-$115325.85 has a best-fit atmosphere solution of $\log{g}=5.82\pm0.02$ and $T_{\text{eff}}$ = 14,780 $\pm$ 70 K (Figure \ref{elm_fits}), making this a clear ELM WD system. We obtained 13 spectra of this object over four nights and detect significant radial velocity variations. However, due to significant period aliasing in the best-fit orbit, further follow-up is required to constrain its orbit and determine companion mass and merger time. TESS full-frame images of J0517$-$1153 do not reveal any significant photometric
variability.

\subsection*{{\rm J0642$-$5605}}

J064207.99$-$560547.44 is an ELM WD with $\log{g}=5.08\pm0.02$ and $T_{\text{eff}}$ = 10,460 $\pm$ 70 K (Figure \ref{elm_fits}). We obtained 14 spectra, resulting in best-fit orbit with period $P=0.13189\pm0.00006$ d and velocity semi-amplitude $K=368.0\pm27.0$ km s$^{-1}$ (Figure \ref{elm_rv_fits}). The minimum companion mass is 0.96 $\pm$ 0.17 M$_\odot$. J0642$-$5605 will merge within 1.3 Gyr.

\begin{figure}[htbp]
\centering
\includegraphics[scale=0.6]{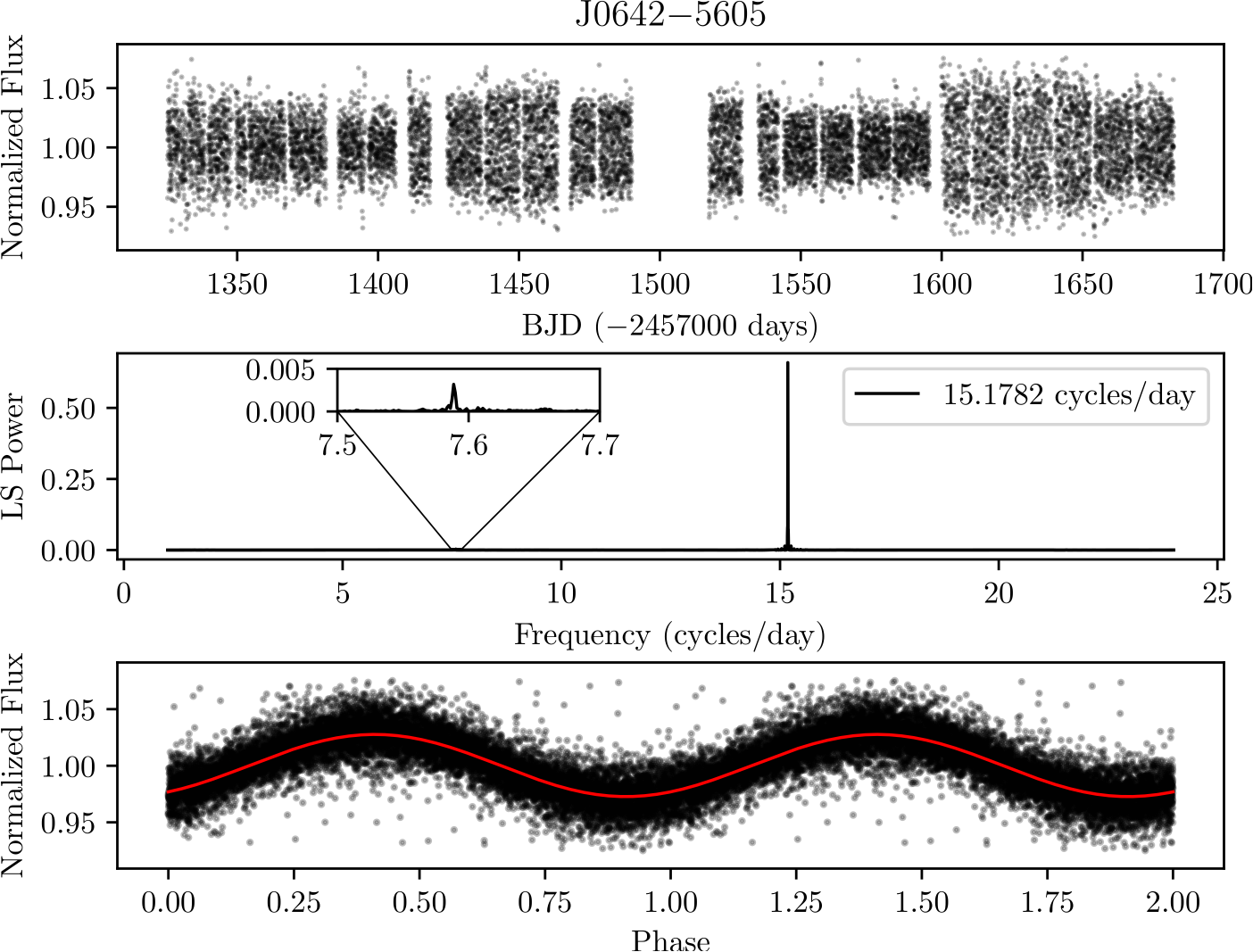}
\caption{TESS Light curve (top), Lomb Scargle periodogram (middle), and phase folded light curve (bottom) of J0642$-$5605. We include a zoomed inset plot showing the region surrounding the small peak at the orbital period of the system. We overplot the best-fit frequency model onto the phase folded light curve for clarity.
}
\label{g138_tess}
\end{figure}

J0642$-$5605 is within the continuous viewing zone of the TESS mission, and was observed as part of Sectors 1-13, except Sector 7.
Figure \ref{g138_tess} shows the TESS FFI light curve of J0642$-$5605 obtained over almost a year, its Lomb Scargle periodogram, and phase folded light curve at the highest-peaked frequency. This WD shows 2.77 $\pm$ 0.02\% photometric variability at a frequency of 15.17820 cycles per day, which is
roughly twice the orbital frequency measured from our radial velocity data. In addition, there is a smaller but significant peak at the
orbital period of the system as well. Hence, TESS data not only confirm the orbital period, but also reveal variability at half the orbital
period, revealing ellipsoidal variations in this system. These variations are intrinsic to the source, and are also confirmed in
the ASAS-SN data \citep{kochanek17}.

\subsection*{{\rm J0650$-$4925}}

J065051.48$-$492549.46 is an ELM WD with best-fit atmosphere solution of $\log{g}=5.47\pm0.03$ and $T_{\text{eff}}$ = 11,210 $\pm$ 90 K (Figure \ref{elm_fits}). From our 13 radial velocity measurements, we obtained a best-fit orbital period $P=0.17453\pm0.00028$ d with velocity semi-amplitude $K=284.2\pm39.4$ km s$^{-1}$ (Figure \ref{elm_rv_fits}). The minimum companion mass is 0.67 $\pm$ 0.21 M$_\odot$. J0650$-$4925 will merge within a Hubble time, with a maximum gravitational wave merger time of 3.6 Gyr.
TESS full-frame images on J0650$-$4925 do not reveal any significant photometric variability.

\subsection*{{\rm J0930$-$8107}}

J093008.47$-$810738.32 is an ELM WD with best-fit atmosphere solution of $\log{g}=6.14\pm0.02$ and $T_{\text{eff}}$ = 23,350 $\pm$ 120 K (Figure \ref{elm_fits}). Fitting 14 radial velocity measurements, we obtained for the best-fit period $P=0.08837\pm0.00005$ d with velocity semi-amplitude $K=212.0\pm9.0$ km s$^{-1}$ (Figure \ref{elm_rv_fits}). J0930$-$8107 has a mass of 0.24 $\pm$ 0.01 M$_\odot$ with minimum companion mass of 0.29 $\pm$ 0.02 M$_\odot$, potentially making this a double ELM WD system. J0930$-$8107 will merge within a Hubble time, with a maximum gravitational wave merger time of 0.9 Gyr.

J0930$-$8107 is included in Sectors 11, 12, and 13 of the TESS mission full-frame images. The combined light curve and its FT
show a peak at 7.084 cycles per day with $0.035 \pm 0.006$ amplitude. However, this peak is only visible in the Sector 11 data,
indicating that it is most likely not intrinsic to the star. J0930$-$8107 is the shortest period system presented here, and the observed
variability in the TESS data does not match the orbital period (11.3 cycles per day), and is likely caused by contamination from
neighboring sources in the TESS images.

\subsection*{{\rm J1239$-$2041}}

J123950.37$-$204142.28 has a best-fit atmosphere solution of $\log{g}=7.03\pm0.04$ and $T_{\text{eff}}$ = 17,750 $\pm$ 210 K (Figure \ref{elm_fits}). We obtained six spectra of J1239$-$2041 over three nights and measure significant radial velocity variations. However, due to significant period aliasing, additional follow-up is required to constrain the orbit and determine companion mass and merger time. Based on the \cite{Istrate2016} He-Core ELM WD models, J1239$-$2041 is a 0.30 $\pm$ 0.01 M$_\odot$ He-core WD.
TESS full-frame images on J1239$-$2041 do not reveal any significant photometric variability.

\subsection{Survey Efficiency}

We observed 82 unique systems using our Gaia parallax target selection method. Of these 82 systems, six contain an ELM WD
based on stellar atmosphere fits. We confirmed all six of these to be in compact binary systems and  obtained precise orbital periods
for four systems, two of which will merge within a Hubble time.

\begin{figure}[htb]
\centering
\vspace{-20pt}
\includegraphics[scale=0.70]{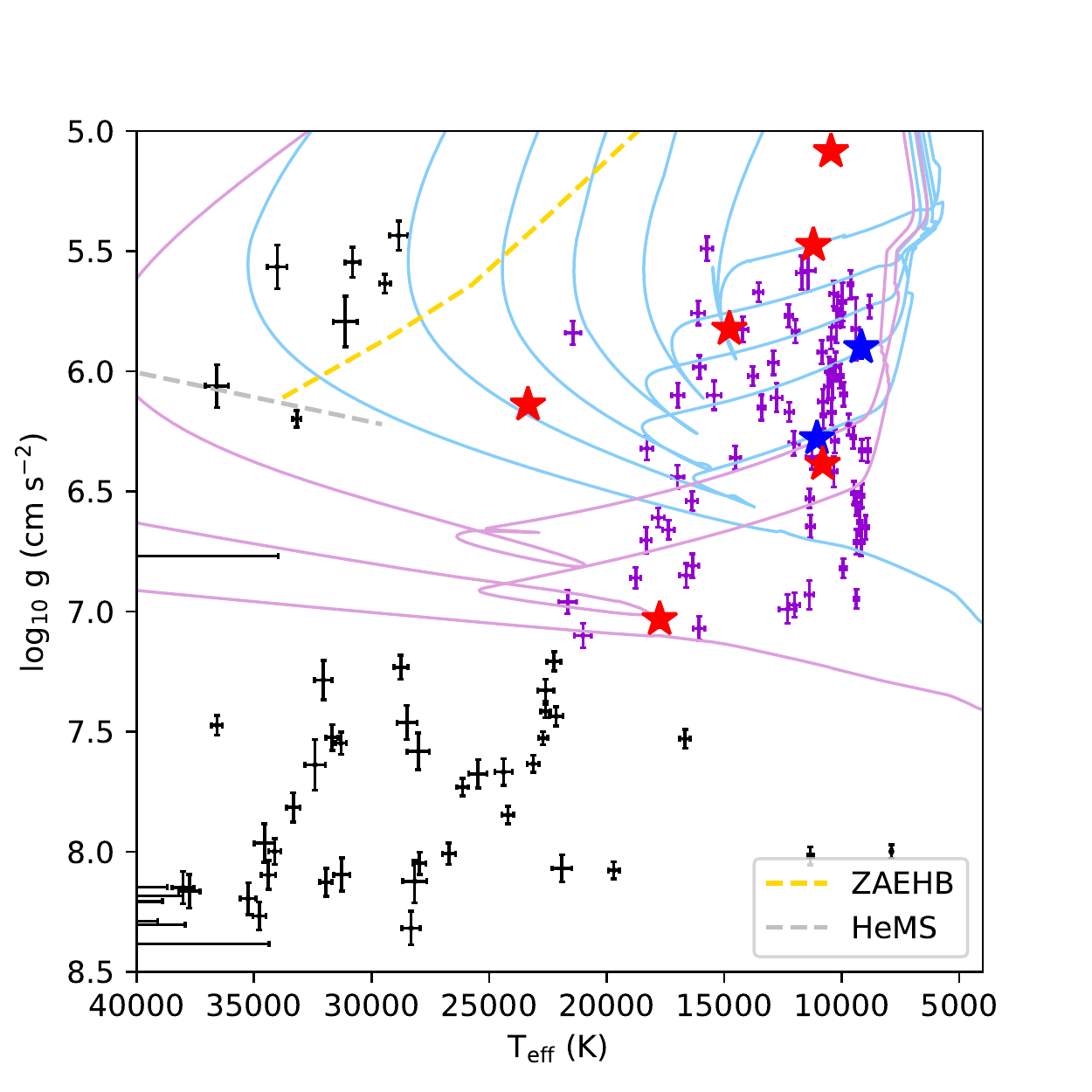}
\caption{log(g) vs T$_{\rm eff}$ plot of the 82 systems observed through our Gaia parallax selection.
Red stars represent of the six ELM systems identified from our Gaia parallax selection.
Blue stars represent two new ELMs identified from our ATLAS+SkyMapper color selection.
Purple points show the locations of previously published ELM WDs from the ELM Survey.
Evolutionary tracks for 0.205 M$_\odot$ (light blue) and 0.306 M$_\odot$ (purple) ELM WDs from Istrate et al. (2016) are overplotted.
Hydrogen shell flashes cause loops seen in the model tracks.
The silver and gold dashed lines show the locations of the Helium main sequence (HeMS) and zero-age extreme-horizontal branch (ZAEHB).}
\label{gaia_teff_plot}
\end{figure}

Figure \ref{gaia_teff_plot} shows a $\log{g}$ vs $T_{\rm eff}$ plot of the objects fit with Hydrogen atmospheres and $\log{g}>5.0$. Black points are objects observed in this survey, identified through Gaia parallax. Red stars mark the location of the six new ELM systems identified through Gaia parallax. Blue stars mark the locations of the two new ELM WDs identified in our ATLAS + SkyMapper color selection discussed earlier in this work. Purple points mark the locations of the ELM WDs previously published in the ELM Survey. We overlay the \citet{Istrate2016} 0.2 M$_\odot$ (light blue) and 0.3 M$_\odot$ (purple) He-core ELM WD evolutionary tracks, Helium main-sequence (HeMS, silver dashed line) and Zero-Age Extreme-Horizontal Branch (ZAEHB, gold dashed line) for reference.

\begin{deluxetable*}{llrrrrrrrrr}[htbp]
\tablecolumns{11}
\tablewidth{0pt}
\tablecaption{The physical parameters of the eight new ELM WDs identified in this work.
Targets marked with a $^\star$ are also included in Pelisoli \& Vos (2019) as ELM WD candidates.}
\tablehead{
\colhead{Gaia Source ID} & \colhead{Object} & \colhead{R.A.} & \colhead{Dec.} & \colhead{Gaia G} & {Gaia Parallax} & \colhead{T$_{\text{eff}}$} & \colhead{log(g)} & \colhead{M$_{\rm {WD}}$} \\
\colhead{} & \colhead{} & \colhead{} & \colhead{} & \colhead{(mag)} & \colhead{(mas)} & \colhead{(K)} & \colhead{(cm s$^{-2}$)} & \colhead{(M$_\odot$)}  }
\startdata
{3680368505418792320} & {J1236$-$0444} & {12:36:19.70} & {$-$04:44:37.90} & {17.29} & {1.91$\pm$0.12} & {11100$\pm$110} & {6.28$\pm$0.02} & {0.156$\pm$0.01} \\
{6308188582700310912} & {J1514$-$1436} & {15:14:47.26} & {$-$14:36:26.77} & {18.27} & {0.57$\pm$0.22} & {9170$\pm$30} & {5.91$\pm$0.05} & {0.167$\pm$0.01} \\
{3183166667278838656} & {J0500$-$0930 $^\star$} & {05:00:51.80} & {$-$09:30:56.98} & {12.62} & {13.97$\pm$0.05} & {10810$\pm$40} & {6.39$\pm$0.02} & {0.163$\pm$0.01} \\
{2989093214186918784} & {J0517$-$1153 $^\star$} & {05:17:24.97} & {$-$11:53:25.85} & {16.22} & {1.56$\pm$0.06} & {14780$\pm$70} & {5.82$\pm$0.02} & {0.179$\pm$0.01} \\
{5496812536854546432} & {J0642$-$5605} $^\star$ & {06:42:07.99} &  {$-$56:05:47.44} & {15.26} & {1.42$\pm$0.03} & {10460$\pm$70} & {5.08$\pm$0.02} & {0.182$\pm$0.01} \\
{5503089133341793792} & {J0650$-$4925 $^\star$} & {06:50:51.48} & {$-$49:25:49.46} & {17.07} & {0.96$\pm$0.06} & {11210$\pm$90} & {5.47$\pm$0.03} & {0.182$\pm$0.01} \\
{5195888264601707392} & {J0930$-$8107 $^\star$} & {09:30:08.47} & {$-$81:07:38.32} & {16.25} & {1.17$\pm$0.04} & {23350$\pm$120} & {6.14$\pm$0.02} & {0.238$\pm$0.01} \\
{3503613283880705664} & {J1239$-$2041} & {12:39:50.37} & {$-$20:41:42.28} & {18.98} & {1.41$\pm$0.33} & {17750$\pm$210} & {7.03$\pm$0.04} & {0.305$\pm$0.01} \\
\enddata
\vspace{-0.8cm}
\end{deluxetable*}
\label{table:elms_table}

In addition to the six new ELM systems, we identify 49 white dwarfs (Table 6), 20 of which are low-mass ($0.3 {\rm M}_\odot \leq M_{\rm WD} \lesssim 0.5 {\rm M}_\odot$), seven subdwarf B stars (Table 7), and four emission-line systems. We present the spectra of the emission line systems in the appendix (Figure \ref{cv}).
We note that 37 of the 49 white dwarfs in Table 6 are hotter than 25,000 K, the upper limit of our target selection criterion. We believe this is due to reddening. Since extinction correction is problematic in Gaia filters, it is not surprising that we are finding a large number of hot WDs contaminating our sample.
The reduced spectra used for atmosphere and orbital fitting for all targets published here is archived in Zenodo\footnote{\url{http://doi.org/10.5281/zenodo.3635104}} in \textmyfont{FITS} format \citep{zenodorepo}.

\section{Summary and Conclusions}

We present the results from a targeted survey for ELM WDs in the southern sky using two different techniques.
Prior to the Gaia DR2, we relied on photometry from the VST ATLAS and SkyMapper surveys to select blue
stars with low-surface gravity.
We note that the VST ATLAS DR4, released April 2019, offers an improved calibration based on Gaia photometry and a larger southern sky footprint over DR2+DR3 used in our survey. Similar to VST ATLAS DR4, SkyMapper DR2 provides not only an extended southern sky footprint, but deeper photometry in the $uvgriz$ bands with limiting magnitudes of about 19 mag in the $g$ and $r$ filters.

With the release of Gaia DR2 astrometry, we developed a new target selection method using Gaia parallax measurements and
tested it in March 2019 using 82 objects. We identified 6 new ELM WD binary systems and 20 additional systems with
$M<0.5 M_{\odot}$, which correspond to $\sim7$\% and $\sim32$\% efficiency for ELM and low-mass WDs with $M<0.5 M_{\odot}$,
respectively. In total, we identified eight new ELM WD systems, and constrained the orbital parameters for six of these systems,
three of which will merge within 4 Gyr. We present a summary of the physical and orbital parameters for these eight
new ELM WD systems in Tables \ref{table:elms_table} and \ref{table:elms_orbit_table}, respectively.

While it appears that Gaia parallax is an efficient method for targeting ELM WDs, we note that \cite{Pelisoli2019b} have created a target list of 5672 (including 2898 with Dec$<0^\circ$) ELM WD candidates based on Gaia colors and astrometry with no restrictions on reddening. Five of our eight new ELM WD systems are also included in \cite{Pelisoli2019b} as ELM WD candidates, but three are missing from their catalog as
\citet{Pelisoli2019b} applied stricter cuts to create their catalog. In addition to these five ELM systems, 27 of our other targets with
SOAR spectra were also included in \cite{Pelisoli2019b}. Almost all of these are normal DA white dwarfs or sdB stars, indicating
a non-negligible contamination of their ELM candidate list. 

\begin{deluxetable*}{lrlrr}[htbp]
\tablecolumns{5}
\tablewidth{0pt}
\tablecaption{Orbital parameters for the six new binaries identified in this
work. Radial velocity measurements for all targets are presented in the Appendix.
Targets marked with a $^\star$ are also included in Pelisoli \& Vos (2019) as ELM WD candidates.}
\tablehead{
\colhead{Object} & \colhead{$P$} & \colhead{$K$} & \colhead{M$_{\text{2,min}}$} & \colhead{$\tau_{\rm max}$ } \\
\colhead{} & \colhead{(d)} & \colhead{(km s$^{-1}$)} & \colhead{(M$_\odot$)} & \colhead{(Gyr)} }
\startdata
{J1236$-$0444} & {0.68758$\pm$0.00327} & {138.0$\pm$6.6} & {0.37$\pm$0.04} & \multicolumn{1}{c}{\nodata} \\
{J1514$-$1436} & {0.58914$\pm$0.00244} & {187.7$\pm$6.6} & {0.63$\pm$0.06} & \multicolumn{1}{c}{\nodata} \\
{J0500$-$0930 $^\star$} & {0.39435$\pm$0.00001} & {146.8$\pm$8.3} & {0.30$\pm$0.04} & \multicolumn{1}{c}{\nodata} \\
{J0642$-$5605 $^\star$} & {0.13189$\pm$0.00006} & {368.0$\pm$27.0} & {0.96$\pm$0.17} & \multicolumn{1}{c}{$1.3$} \\
{J0650$-$4925 $^\star$} & {0.17453$\pm$0.00028} & {284.2$\pm$39.4} & {0.67$\pm$0.21} & \multicolumn{1}{c}{$3.6$} \\
{J0930$-$8107 $^\star$} & {0.08837$\pm$0.00005} & {212.0$\pm$9.0} & {0.29$\pm$0.03} & \multicolumn{1}{c}{$0.9$} \\
\enddata
\vspace{-0.8cm}
\end{deluxetable*}
\label{table:elms_orbit_table}

The shortest period ELM WD binaries will serve as multi-messenger laboratories, when they are detected by the Laser Interferometer
Space Antenna (LISA). Hence, the discovery of additional systems now is important for characterizing such systems before LISA is
operational. We are continuing to observe the remaining Gaia selected targets in our sample, and along with the eclipsing and/or
tidally distorted ELM WD discoveries from the Zwicky Transient Facility \citep{Burdge2019b} and the upcoming Large Synoptic Survey
Telescope, we hope to significantly increase the known population of ELM WDs in the next few years.

\acknowledgements
We thank the anonymous referee for helpful comments and suggestions that greatly improved the quality of this work. This work was supported in part by the Smithsonian Institution, and in part by the NSF under grant AST-1906379.
This project makes use of data obtained at the Southern Astrophysical Research (SOAR) telescope, which is a
joint project of the Minist\'{e}rio da Ci\^{e}ncia, Tecnologia, Inova\c{c}\~{a}os e
Comunica\c{c}\~{a}oes do Brasil, the U.S.\ National Optical Astronomy Observatory,
the University of North Carolina at Chapel Hill, and Michigan State University.
This research made use of Astropy,\footnote{http://www.astropy.org} a community-developed core Python package for Astronomy \citep{astropy2013, astropy2018}. 

\facilities{MMT (Blue Channel spectrograph), FLWO:1.5m (FAST spectrograph), SOAR (Goodman spectrograph), Magellan (MagE)}

\appendix

\section{Additional Systems: Emission Line Objects}

\begin{figure}[htbp]
\centering
\includegraphics[scale=0.2,angle=0]{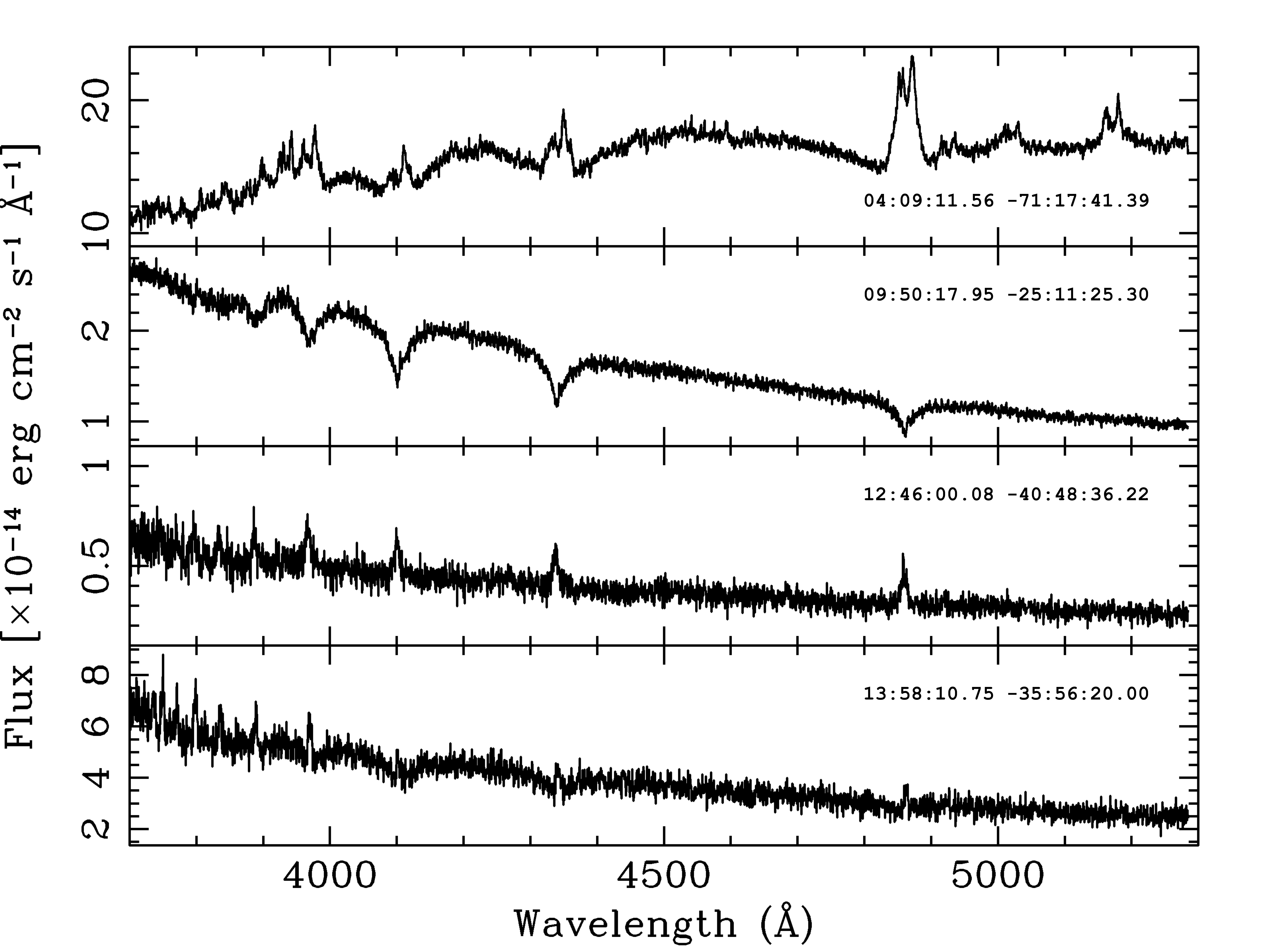}
\caption{Spectra of emission line objects observed as a part of the ELM Survey South: I.
Hydrogen emission lines can be seen in the core of broad Balmer lines in all objects.
J0409$-$7117 shows multi-peaked Hydrogen and Metal emission lines.
J0950$-$2511 shows faint emission lines systematically offset towards the red wing of Balmer line cores.
Both J0409$-$7117 and J1358$-$3556 were identified as CV or WD+M candidates by \citet{Pelisoli2019b}.}
\label{cv}
\end{figure}

Among all of the systems observed throughout our survey, we identified a handful of emission line systems. For completeness, here
we display the optical spectrum for these four objects (Figure \ref{cv}). J0409$-$7117 (Figure \ref{cv}, top) shows evidence of an accretion disk in its Balmer and metal (e.g., Mg) emission lines. J0409$-$7117 was identified as a CV or WD+M candidate by \citet{Pelisoli2019b}.
One of these emission line objects, J1358$-$3556 (Figure \ref{cv}, bottom), shows variability at a frequency of 12.3 cycles per day in the TESS full-frame images. J1358$-$3556 was also identified as a CV or WD+M candidate by \citet{Pelisoli2019b}. There are two additional targets in our sample that show variability in TESS data. J0950$-$2511 is a low-mass WD with an estimated mass of $M=0.44 \pm 0.02 M_{\odot}$, but with weak Balmer emission lines visible in the line cores. 
The Catalina Sky Survey found variations with a period of 0.318654 d \citep{drake17}, and TESS full-frame images also show variability
at the same period. In addition, J0711$-$6727 shows significant variations at a frequency of 4.86 cycles per day.
Follow-up spectroscopy would be useful to constrain the nature of variability in these systems.

\newpage

\section{Data Tables}
\startlongtable
\begin{deluxetable*}{llrrrrrrr}
\tablecolumns{9}
\tablewidth{0pt}
\tablecaption{Table of 123 white dwarfs observed as a part of our ATLAS+SkyMapper target selection.
Targets marked with a $^\star$ are also included in Pelisoli \& Vos (2019) as ELM WD candidates.}
\tablehead{
\colhead{Gaia Source ID} & \colhead{Object} & \colhead{R.A.} & \colhead{Dec.} & \colhead{Gaia G} & \colhead{Gaia Parallax} & \colhead{T$_{\text{eff}}$} & \colhead{log(g)} & \colhead{M$_{\rm WD}$} \\
\colhead{} & \colhead{} & \colhead{} & \colhead{} & \colhead{(mag)} & \colhead{(mas)} & \colhead{(K)} & \colhead{(cm s$^{-2}$)} & \colhead{(M$_\odot$)} }
\startdata
{2314720431736648960} & {J0000$-$3102} & {00:00:41.06} & {$-$31:02:45.82} & {20.05} & {2.95$\pm$1.43} & {13810$\pm$1950} & {7.90$\pm$0.16} & 	{0.55$\pm$0.09} \\
{2421224556841719680} & {J0001$-$1218} & {00:01:17.04} & {$-$12:18:42.66} & {20.09} & {2.28$\pm$0.86} & {12680$\pm$1660} & {8.98$\pm$0.21} & 	{1.19$\pm$0.09} \\
{2427460643197388672} & {J0011$-$1143} & {00:11:04.66} & {$-$11:43:49.98} & {17.08} & {0.25$\pm$0.11} & {11910$\pm$330} & {8.25$\pm$0.08} & 	{0.76$\pm$0.06} \\
{2316638774585068672} & {J0021$-$3154} & {00:21:40.44} & {$-$31:54:42.44} & {18.53} & {0.73$\pm$0.27} & {11250$\pm$380} & {8.02$\pm$0.14} & 	{0.62$\pm$0.09} \\
{2318640465567548800} & {J0024$-$2933} & {00:24:29.67} & {$-$29:33:38.45} & {19.88} & {3.13$\pm$0.78} & {10030$\pm$190} & {7.95$\pm$0.16} & 	{0.57$\pm$0.09} \\
{2424786459119647104} & {J0024$-$1107} & {00:24:54.96} & {$-$11:07:43.28} & {19.46} & {2.86$\pm$0.62} & {13510$\pm$910} & {7.96$\pm$0.12} & 	{0.59$\pm$0.07} \\
{2315815721412502656} & {J0026$-$3224} & {00:26:06.30} & {$-$32:24:23.77} & {17.05} & {9.09$\pm$0.10} & {13160$\pm$810} & {8.42$\pm$0.11} & 	{0.87$\pm$0.08} \\
{2319211352620639616} & {J0030$-$2803} & {00:30:53.20} & {$-$28:03:36.11} & {19.90} & {2.31$\pm$0.63} & {12010$\pm$450} & {8.69$\pm$0.09} & 	{1.04$\pm$0.06} \\
{2346428148058537856} & {J0035$-$2627} & {00:35:49.84} & {$-$26:27:19.84} & {19.55} & {2.18$\pm$0.49} & {11490$\pm$420} & {8.33$\pm$0.12} & 	{0.81$\pm$0.08} \\
{2370382902950807296} & {J0036$-$1657} & {00:36:25.85} & {$-$16:57:18.50} & {15.79} & {0.17$\pm$0.08} & {9940$\pm$200} & {7.12$\pm$0.28} & 	{0.28$\pm$0.07} \\
{2343867935233173376} & {J0041$-$2609} & {00:41:18.62} & {$-$26:09:11.88} & {19.88} & {2.73$\pm$0.61} & {13640$\pm$770} & {8.11$\pm$0.11} & 	{0.68$\pm$0.07} \\
{5005923029327350656} & {J0047$-$3425} & {00:47:28.14} & {$-$34:25:35.47} & {19.64} & {2.49$\pm$0.47} & {13600$\pm$1850} & {7.84$\pm$0.17} & 	{0.52$\pm$0.09} \\
{2474049699645450368} & {J0049$-$0913} & {00:49:44.65} & {$-$09:13:14.23} & {20.30} & {1.53$\pm$0.81} & {15350$\pm$3660} & {8.50$\pm$0.21} & 	{0.93$\pm$0.13} \\
{2473843786029439104} & {J0052$-$0924} & {00:52:15.36} & {$-$09:24:18.71} & {15.42} & {15.69$\pm$0.06} & {13930$\pm$1950} & {7.95$\pm$0.16} & 	{0.58$\pm$0.09} \\
{2473096358640407168} & {J0102$-$1015} & {01:02:16.05} & {$-$10:15:04.18} & {16.76} & {7.08$\pm$0.08} & {18520$\pm$1010} & {7.74$\pm$0.18} & 	{0.49$\pm$0.09} \\
{2456134364556362624} & {J0112$-$1338} & {01:12:15.54} & {$-$13:38:02.87} & {20.17} & {0.70$\pm$0.92} & {26070$\pm$630} & {7.55$\pm$0.09} & 	{0.46$\pm$0.03} \\
{5029203259605424512} & {J0119$-$3002} & {01:19:33.67} & {$-$30:02:02.62} & {19.31} & {1.31$\pm$0.39} & {13100$\pm$980} & {7.74$\pm$0.16} & 	{0.47$\pm$0.08} \\
{5014943044766189312} & {J0134$-$3422} & {01:34:59.92} & {$-$34:22:31.87} & {19.30} & {4.48$\pm$0.32} & {11840$\pm$750} & {8.41$\pm$0.17} & 	{0.86$\pm$0.11} \\
{4969151267391273728} & {J0158$-$3430} & {01:58:11.12} & {$-$34:30:40.50} & {19.41} & {2.34$\pm$0.37} & {14990$\pm$2200} & {8.41$\pm$0.14} & 	{0.87$\pm$0.09} \\
{2461668030485282432} & {J0204$-$1024} & {02:04:10.06} & {$-$10:24:36.72} & {19.60} & {2.11$\pm$0.62} & {11430$\pm$440} & {7.84$\pm$0.18} & 	{0.52$\pm$0.10} \\
{4969649621036866688} & {J0205$-$3319} & {02:05:10.90} & {$-$33:19:16.61} & {19.69} & {2.36$\pm$0.38} & {13770$\pm$1500} & {8.04$\pm$0.17} & 	{0.63$\pm$0.10} \\
{4970933580036041216} & {J0208$-$3328 $^\star$} & {02:08:13.49} & {$-$33:28:11.46} & {18.92} & {1.56$\pm$0.25} & {13530$\pm$1420} & {7.87$\pm$0.15} & 	{0.54$\pm$0.08} \\
{4970170445950131456} & {J0214$-$3344} & {02:14:21.85} & {$-$33:44:21.44} & {18.31} & {$-$0.13$\pm$0.16} & {12400$\pm$440} & {8.18$\pm$0.10} & 	{0.72$\pm$0.07} \\
{5150847728544652416} & {J0217$-$1133} & {02:17:43.89} & {$-$11:33:08.15} & {17.09} & {9.67$\pm$0.11} & {15120$\pm$2670} & {8.55$\pm$0.17} & 	{0.96$\pm$0.11} \\
{5176077603391562752} & {J0217$-$0939} & {02:17:45.97} & {$-$09:39:53.86} & {19.34} & {3.18$\pm$0.64} & {13790$\pm$1980} & {8.08$\pm$0.23} & 	{0.66$\pm$0.14} \\
{4970619149773451136} & {J0219$-$3239} & {02:19:51.61} & {$-$32:39:55.37} & {20.01} & {1.73$\pm$0.55} & {14420$\pm$1770} & {8.18$\pm$0.12} & 	{0.72$\pm$0.08} \\
{5066933619588166400} & {J0221$-$3113} & {02:21:28.37} & {$-$31:13:10.52} & {19.80} & {1.47$\pm$0.44} & {15940$\pm$750} & {8.13$\pm$0.12} & 	{0.69$\pm$0.08} \\
{5050702285341454336} & {J0249$-$3328} & {02:49:58.92} & {$-$33:28:30.22} & {20.01} & {2.79$\pm$0.50} & {11930$\pm$400} & {8.02$\pm$0.12} & 	{0.62$\pm$0.08} \\
{5072652523161819776} & {J0256$-$2632} & {02:56:39.12} & {$-$26:32:34.55} & {18.15} & {0.37$\pm$0.17} & {14050$\pm$950} & {7.77$\pm$0.12} & 	{0.49$\pm$0.06} \\
{5165709277461403520} & {J0310$-$1129} & {03:10:27.67} & {$-$11:29:15.72} & {20.18} & {2.23$\pm$0.89} & {13600$\pm$1030} & {7.42$\pm$0.19} & 	{0.38$\pm$0.07} \\
{5073564636776148992} & {J0314$-$2539} & {03:14:30.95} & {$-$25:39:27.29} & {19.88} & {1.72$\pm$0.46} & {13800$\pm$1610} & {7.98$\pm$0.15} & 	{0.60$\pm$0.09} \\
{5156216368944753024} & {J0314$-$1326} & {03:14:33.00} & {$-$13:26:58.88} & {20.29} & {1.12$\pm$1.27} & {12780$\pm$1240} & {7.85$\pm$0.32} & 	{0.52$\pm$0.17} \\
{5053174743394426496} & {J0322$-$3422} & {03:22:57.76} & {$-$34:22:00.80} & {19.82} & {1.52$\pm$0.45} & {13780$\pm$720} & {8.47$\pm$0.09} & 	{0.91$\pm$0.06} \\
{5161187535892016768} & {J0324$-$1346} & {03:24:09.35} & {$-$13:46:24.31} & {19.74} & {$-$3.25$\pm$1.11} & {15130$\pm$810} & {7.65$\pm$0.16} & 	{0.45$\pm$0.07} \\
{5060865792871440384} & {J0327$-$2632} & {03:27:41.16} & {$-$26:32:31.56} & {20.22} & {1.89$\pm$0.61} & {11270$\pm$400} & {8.25$\pm$0.14} & 	{0.76$\pm$0.09} \\
{4863642926160372608} & {J0343$-$3137} & {03:43:25.14} & {$-$31:37:35.65} & {19.89} & {0.84$\pm$0.47} & {11920$\pm$480} & {7.57$\pm$0.22} & 	{0.41$\pm$0.09} \\
{5085310758152519680} & {J0343$-$2442} & {03:43:28.21} & {$-$24:42:43.52} & {19.44} & {2.22$\pm$0.36} & {12480$\pm$410} & {7.87$\pm$0.11} & 	{0.53$\pm$0.06} \\
{5114162179486492800} & {J0343$-$1318} & {03:43:28.70} & {$-$13:18:16.00} & {16.33} & {10.61$\pm$0.08} & {13780$\pm$920} & {8.03$\pm$0.10} & 	{0.63$\pm$0.06} \\
{3194415427145350656} & {J0351$-$0927} & {03:51:24.91} & {$-$09:27:42.26} & {19.65} & {1.86$\pm$0.63} & {18450$\pm$830} & {7.70$\pm$0.15} & 	{0.47$\pm$0.07} \\
{3193341857120018304} & {J0353$-$1110} & {03:53:26.14} & {$-$11:10:47.39} & {16.08} & {8.29$\pm$0.07} & {18010$\pm$880} & {8.03$\pm$0.15} & 	{0.64$\pm$0.09} \\
{5093664194662891776} & {J0353$-$2123} & {03:53:59.40} & {$-$21:23:21.52} & {19.51} & {1.23$\pm$0.43} & {13360$\pm$1120} & {7.76$\pm$0.12} & 	{0.48$\pm$0.06} \\
{4886897352272683008} & {J0357$-$3026} & {03:57:21.79} & {$-$30:26:30.73} & {19.89} & {2.69$\pm$0.44} & {14390$\pm$1940} & {8.13$\pm$0.14} & 	{0.69$\pm$0.09} \\
{4883110428068169472} & {J0401$-$3332} & {04:01:17.81} & {$-$33:32:05.28} & {19.18} & {2.92$\pm$0.26} & {12740$\pm$500} & {7.79$\pm$0.14} & 	{0.49$\pm$0.07} \\
{5686477223197435776} & {J1002$-$1445} & {10:02:27.13} & {$-$14:45:37.76} & {19.41} & {3.73$\pm$0.96} & {14080$\pm$1310} & {8.03$\pm$0.10} & 	{0.63$\pm$0.06} \\
{5686544568284108416} & {J1007$-$1430} & {10:07:17.01} & {$-$14:30:58.36} & {20.40} & {3.17$\pm$0.90} & {15300$\pm$1090} & {8.27$\pm$0.10} & 	{0.78$\pm$0.07} \\
{3768573016120952832} & {J1017$-$0940} & {10:17:56.48} & {$-$09:40:18.41} & {19.59} & {3.06$\pm$0.79} & {12450$\pm$670} & {8.22$\pm$0.14} & 	{0.74$\pm$0.09} \\
{3779023285732532736} & {J1018$-$0524} & {10:18:12.22} & {$-$05:24:47.45} & {19.86} & {2.56$\pm$1.15} & {14700$\pm$1030} & {8.02$\pm$0.12} & 	{0.62$\pm$0.07} \\
{3768136711868126208} & {J1021$-$0946} & {10:21:12.12} & {$-$09:46:09.66} & {20.30} & \multicolumn{1}{c}{\nodata} & {11040$\pm$280} & {8.26$\pm$0.11} & 	{0.77$\pm$0.08} \\
{3768242505503031552} & {J1021$-$0908} & {10:21:22.59} & {$-$09:08:38.33} & {18.83} & {3.75$\pm$0.40} & {13830$\pm$710} & {8.01$\pm$0.07} & 	{0.62$\pm$0.05} \\
{3861429723729285376} & {J1021$+$0543} & {10:21:53.13} & {$+$05:43:22.37} & {19.67} & {1.36$\pm$0.60} & {9600$\pm$50} & {7.54$\pm$0.04} & 	{0.40$\pm$0.02} \\
{3769160528992333824} & {J1022$-$0808} & {10:22:59.07} & {$-$08:08:20.22} & {19.80} & {4.49$\pm$1.77} & {13030$\pm$750} & {8.21$\pm$0.13} & 	{0.74$\pm$0.09} \\
{3781985683590564736} & {J1023$-$0251} & {10:23:39.00} & {$-$02:51:23.24} & {17.07} & {3.27$\pm$0.16} & {30630$\pm$1170} & {7.54$\pm$0.22} & 	{0.46$\pm$0.08} \\
{3749062659027515904} & {J1026$-$1538} & {10:26:02.02} & {$-$15:38:38.29} & {20.29} & \multicolumn{1}{c}{\nodata} & {14330$\pm$1350} & {8.20$\pm$0.10} & 	{0.73$\pm$0.07} \\
{3755961063699688448} & {J1029$-$1002} & {10:29:10.41} & {$-$10:02:47.83} & {19.70} & {3.33$\pm$0.59} & {13490$\pm$1570} & {7.81$\pm$0.15} & 	{0.51$\pm$0.08} \\
{3748932298180213504} & {J1033$-$1537} & {10:33:08.66} & {$-$15:37:14.20} & {18.23} & {3.29$\pm$0.26} & {13380$\pm$1310} & {7.76$\pm$0.14} & 	{0.48$\pm$0.07} \\
{3781482309128453760} & {J1038$-$0248} & {10:38:32.71} & {$-$02:48:41.30} & {17.22} & {6.66$\pm$0.15} & {14110$\pm$1040} & {8.07$\pm$0.08} & 	{0.65$\pm$0.05} \\
{3762701108633074688} & {J1040$-$0746 $^\star$} & {10:40:26.09} & {$-$07:46:14.97} & {17.76} & {2.16$\pm$0.16} & {24670$\pm$570} & {7.60$\pm$0.09} & 	{0.47$\pm$0.04} \\
{3802511018765753728} & {J1045$-$0212} & {10:45:54.42} & {$-$02:12:29.00} & {19.19} & {1.98$\pm$0.48} & {14140$\pm$1240} & {8.29$\pm$0.13} & 	{0.79$\pm$0.09} \\
{3761504285930683008} & {J1047$-$0919} & {10:47:19.75} & {$-$09:19:00.01} & {19.63} & {1.97$\pm$0.60} & {11870$\pm$350} & {8.04$\pm$0.10} & 	{0.63$\pm$0.07} \\
{3762845930634843264} & {J1050$-$0905} & {10:50:21.96} & {$-$09:05:11.11} & {17.81} & {4.99$\pm$0.17} & {14570$\pm$710} & {8.01$\pm$0.08} & 	{0.62$\pm$0.05} \\
{3760440267912519936} & {J1102$-$0757} & {11:02:11.77} & {$-$07:57:53.46} & {17.92} & {2.93$\pm$0.22} & {28380$\pm$560} & {8.15$\pm$0.10} & 	{0.73$\pm$0.06} \\
{3787385900590991232} & {J1123$-$0302} & {11:23:18.74} & {$-$03:02:48.88} & {17.26} & {6.68$\pm$0.24} & {13970$\pm$1440} & {8.11$\pm$0.14} & 	{0.68$\pm$0.09} \\
{3591995876813375232} & {J1132$-$0822} & {11:32:09.43} & {$-$08:22:47.89} & {18.40} & {4.90$\pm$0.30} & {12740$\pm$390} & {7.99$\pm$0.09} & 	{0.60$\pm$0.06} \\
{3793382563994044416} & {J1133$-$0222} & {11:33:23.29} & {$-$02:22:28.64} & {20.40} & {0.02$\pm$1.35} & {11000$\pm$410} & {8.02$\pm$0.18} & 	{0.62$\pm$0.11} \\
{3593234201782795008} & {J1133$-$0624} & {11:33:26.97} & {$-$06:24:26.24} & {20.36} & {2.30$\pm$1.23} & {13980$\pm$1270} & {8.07$\pm$0.11} & 	{0.65$\pm$0.07} \\
{3586969459405663744} & {J1134$-$1023} & {11:34:47.03} & {$-$10:23:02.26} & {19.68} & {1.73$\pm$0.62} & {16000$\pm$430} & {7.86$\pm$0.08} & 	{0.54$\pm$0.05} \\
{3560803316046802432} & {J1134$-$1454} & {11:34:57.05} & {$-$14:54:59.54} & {20.20} & {0.75$\pm$0.83} & {11470$\pm$290} & {8.28$\pm$0.09} & 	{0.78$\pm$0.06} \\
{3593790318443372288} & {J1136$-$0523} & {11:37:36.90} & {$-$05:23:16.67} & {16.81} & {6.65$\pm$0.12} & {19600$\pm$610} & {8.20$\pm$0.09} & 	{0.74$\pm$0.06} \\
{3595269775762905472} & {J1156$-$0648} & {11:56:31.41} & {$-$06:48:20.80} & {18.05} & {3.76$\pm$0.19} & {13310$\pm$2020} & {7.68$\pm$0.20} & 	{0.45$\pm$0.09} \\
{3594535198915846784} & {J1204$-$0737} & {12:04:48.44} & {$-$07:37:21.07} & {19.41} & {2.98$\pm$0.41} & {12480$\pm$520} & {8.08$\pm$0.12} & 	{0.65$\pm$0.08} \\
{3582049385389446912} & {J1205$-$0826} & {12:05:45.49} & {$-$08:26:41.96} & {19.81} & {2.62$\pm$0.51} & {13280$\pm$500} & {8.09$\pm$0.09} & 	{0.66$\pm$0.06} \\
{3581918887102403200} & {J1206$-$0924} & {12:06:13.70} & {$-$09:24:42.52} & {19.09} & {2.48$\pm$0.33} & {11590$\pm$190} & {7.77$\pm$0.08} & 	{0.48$\pm$0.05} \\
{3596970101775019392} & {J1213$-$0518} & {12:13:49.16} & {$-$05:18:56.50} & {18.26} & {4.05$\pm$0.21} & {14190$\pm$990} & {8.25$\pm$0.10} & 	{0.76$\pm$0.07} \\
{3682256405538928512} & {J1250$-$0330} & {12:50:49.69} & {$-$03:30:20.91} & {19.57} & {3.23$\pm$0.43} & {12160$\pm$380} & {8.01$\pm$0.10} & 	{0.61$\pm$0.06} \\
{3682458814461982592} & {J1254$-$0218} & {12:54:58.08} & {$-$02:18:37.77} & {16.65} & {8.14$\pm$0.12} & {13410$\pm$380} & {8.30$\pm$0.06} & 	{0.80$\pm$0.05} \\
{3635681505303143296} & {J1316$-$0510} & {13:16:27.72} & {$-$05:10:26.00} & {15.83} & {0.09$\pm$0.08} & {13790$\pm$1130} & {8.37$\pm$0.15} & 	{0.84$\pm$0.10} \\
{3656994674918366208} & {J1350$-$0424} & {13:50:53.32} & {$-$04:24:36.86} & {19.72} & {3.03$\pm$0.63} & {12560$\pm$430} & {8.17$\pm$0.09} & 	{0.71$\pm$0.06} \\
{3645049580594224512} & {J1355$-$0415} & {13:55:45.55} & {$-$04:15:21.26} & {19.66} & {4.09$\pm$0.95} & {13560$\pm$1620} & {7.95$\pm$0.20} & 	{0.58$\pm$0.11} \\
{6311053875642748928} & {J1438$-$1418} & {14:38:53.93} & {$-$14:18:21.17} & {19.28} & {3.23$\pm$0.42} & {15960$\pm$520} & {8.17$\pm$0.08} & 	{0.72$\pm$0.06} \\
{6281939426215247872} & {J1444$-$1827} & {14:44:03.97} & {$-$18:27:09.58} & {20.08} & {2.64$\pm$0.81} & {15270$\pm$890} & {8.03$\pm$0.13} & 	{0.63$\pm$0.08} \\
{6337488780874439680} & {J1450$-$0451} & {14:50:50.24} & {$-$04:51:35.78} & {20.00} & \multicolumn{1}{c}{\nodata} & {13020$\pm$690} & {8.30$\pm$0.11} & 	{0.79$\pm$0.08} \\
{6310239721642756224} & {J1454$-$1440} & {14:54:21.85} & {$-$14:40:29.24} & {19.98} & {3.99$\pm$0.92} & {11550$\pm$330} & {8.25$\pm$0.09} & 	{0.76$\pm$0.06} \\
{6310532019936081536} & {J1454$-$1418} & {14:54:27.96} & {$-$14:18:16.67} & {20.16} & {1.72$\pm$1.36} & {11650$\pm$310} & {7.95$\pm$0.11} & 	{0.58$\pm$0.07} \\
{6333254531531385088} & {J1455$-$0736} & {14:55:10.25} & {$-$07:36:44.24} & {19.30} & {2.82$\pm$0.50} & {12280$\pm$310} & {8.00$\pm$0.08} & 	{0.61$\pm$0.05} \\
{6337391027418990464} & {J1457$-$0423} & {14:57:49.40} & {$-$04:23:07.19} & {19.56} & {0.39$\pm$0.49} & {8750$\pm$230} & {8.53$\pm$0.24} & 	{0.94$\pm$0.15} \\
{6310291398688877312} & {J1459$-$1433} & {14:59:28.15} & {$-$14:33:26.10} & {20.05} & \multicolumn{1}{c}{\nodata} & {14740$\pm$1090} & {8.06$\pm$0.11} & 	{0.65$\pm$0.07} \\
{6257011607827666560} & {J1501$-$1913} & {15:01:37.26} & {$-$19:13:09.59} & {19.21} & {2.86$\pm$0.37} & {12030$\pm$400} & {7.91$\pm$0.12} & 	{0.56$\pm$0.07} \\
{6305223680876589952} & {J1504$-$1816} & {15:04:17.22} & {$-$18:16:23.05} & {19.71} & {2.58$\pm$0.48} & {11260$\pm$290} & {7.86$\pm$0.12} & 	{0.53$\pm$0.07} \\
{6322910390561746560} & {J1519$-$0530} & {15:19:19.89} & {$-$05:30:56.84} & {19.98} & {2.84$\pm$0.92} & {14110$\pm$820} & {7.71$\pm$0.12} & 	{0.46$\pm$0.06} \\
{6258275943120737792} & {J1520$-$1725} & {15:20:47.27} & {$-$17:25:23.48} & {19.42} & {1.86$\pm$0.50} & {19030$\pm$700} & {7.91$\pm$0.12} & 	{0.57$\pm$0.06} \\
{6322818753139011712} & {J1522$-$0600} & {15:22:18.42} & {$-$06:00:50.72} & {20.13} & {2.24$\pm$0.73} & {11520$\pm$300} & {7.51$\pm$0.15} & 	{0.39$\pm$0.06} \\
{6321330392351972992} & {J1522$-$0743} & {15:22:40.59} & {$-$07:43:49.19} & {20.18} & \multicolumn{1}{c}{\nodata} & {14710$\pm$1720} & {8.50$\pm$0.15} & 	{0.93$\pm$0.10} \\
{6266081754122219776} & {J1528$-$1407} & {15:28:13.98} & {$-$14:07:58.19} & {20.49} & {$-$0.28$\pm$1.21} & {13220$\pm$790} & {8.33$\pm$0.11} & 	{0.81$\pm$0.08} \\
{6842569307021815552} & {J2129$-$1357} & {21:29:02.60} & {$-$13:57:59.00} & {16.21} & {7.24$\pm$0.12} & {24180$\pm$620} & {8.12$\pm$0.11} & 	{0.70$\pm$0.07} \\
{6810587606146932096} & {J2138$-$2822} & {21:38:22.55} & {$-$28:22:01.27} & {20.16} & {$-$0.82$\pm$1.16} & {11370$\pm$620} & {8.11$\pm$0.21} & 	{0.67$\pm$0.13} \\
{6816647220885100928} & {J2145$-$2256} & {21:45:12.49} & {$-$22:56:22.74} & {18.04} & {0.02$\pm$0.22} & {13240$\pm$850} & {7.99$\pm$0.16} & 	{0.60$\pm$0.10} \\
{6617354691037040256} & {J2149$-$2945} & {21:49:47.26} & {$-$29:45:43.31} & {17.62} & {0.15$\pm$0.14} & {13120$\pm$740} & {7.91$\pm$0.15} & 	{0.56$\pm$0.09} \\
{6588218560614027392} & {J2156$-$3515} & {21:56:37.62} & {$-$35:15:47.34} & {20.15} & {2.16$\pm$1.09} & {11400$\pm$560} & {8.87$\pm$0.16} & 	{1.14$\pm$0.08} \\
{6812758591855108736} & {J2200$-$2400} & {22:00:22.24} & {$-$24:00:21.17} & {17.88} & {$-$0.04$\pm$0.20} & {12320$\pm$1170} & {8.62$\pm$0.20} & 	{1.00$\pm$0.12} \\
{6573507576070298880} & {J2206$-$3929} & {22:06:59.42} & {$-$39:29:47.54} & {19.80} & {1.71$\pm$0.79} & {14300$\pm$1500} & {8.21$\pm$0.12} & 	{0.74$\pm$0.08} \\
{6619862642700459904} & {J2207$-$2611} & {22:07:52.86} & {$-$26:11:09.90} & {19.36} & {3.17$\pm$0.50} & {11390$\pm$670} & {8.43$\pm$0.20} & 	{0.88$\pm$0.13} \\
{6618244814419373184} & {J2212$-$2828} & {22:12:14.57} & {$-$28:28:49.40} & {17.08} & {0.12$\pm$0.11} & {18060$\pm$2250} & {7.64$\pm$0.42} & 	{0.45$\pm$0.18} \\
{2601267970982861952} & {J2230$-$1250} & {22:30:28.84} & {$-$12:50:51.32} & {20.34} & {3.08$\pm$1.98} & {13300$\pm$900} & {7.83$\pm$0.13} & 	{0.52$\pm$0.07} \\
{6600835284744283520} & {J2238$-$3241} & {22:38:43.07} & {$-$32:41:28.54} & {18.36} & {0.04$\pm$0.25} & {10970$\pm$410} & {8.51$\pm$0.16} & 	{0.93$\pm$0.11} \\
{6623866514292772352} & {J2239$-$2436} & {22:39:30.02} & {$-$24:36:39.38} & {17.77} & {0.12$\pm$0.21} & {10480$\pm$300} & {7.89$\pm$0.19} & 	{0.54$\pm$0.11} \\
{6604376158861666560} & {J2248$-$3103} & {22:48:34.39} & {$-$31:03:06.48} & {19.52} & {2.57$\pm$0.63} & {14200$\pm$2750} & {7.97$\pm$0.24} & 	{0.59$\pm$0.14} \\
{6548178264140855424} & {J2251$-$3624} & {22:51:14.68} & {$-$36:24:20.02} & {19.22} & {1.69$\pm$0.38} & {12780$\pm$660} & {7.58$\pm$0.14} & 	{0.42$\pm$0.06} \\
{6609954810278528128} & {J2254$-$2721} & {22:54:03.94} & {$-$27:21:15.59} & {19.76} & {2.55$\pm$0.64} & {15440$\pm$620} & {7.17$\pm$0.14} & 	{0.31$\pm$0.03} \\
{2603274858876574720} & {J2256$-$1319} & {22:56:13.00} & {$-$13:19:39.00} & {15.68} & {10.48$\pm$0.09} & {20690$\pm$530} & {7.90$\pm$0.08} & 	{0.57$\pm$0.05} \\
{2396582989434764672} & {J2256$-$2224} & {22:56:44.93} & {$-$22:24:33.41} & {20.31} & {2.39$\pm$0.81} & {12360$\pm$430} & {8.06$\pm$0.11} & 	{0.64$\pm$0.07} \\
{2383060130285253632} & {J2307$-$2518} & {23:07:02.01} & {$-$25:18:05.90} & {18.75} & {$-$0.33$\pm$0.33} & {10800$\pm$380} & {8.66$\pm$0.17} & 	{1.02$\pm$0.10} \\
{2379359556397873664} & {J2308$-$2710} & {23:08:52.37} & {$-$27:10:29.21} & {16.07} & {6.24$\pm$0.08} & {17400$\pm$640} & {7.41$\pm$0.13} & 	{0.39$\pm$0.04} \\
{6556568740451265152} & {J2309$-$3217} & {23:09:14.85} & {$-$32:17:18.31} & {19.47} & {2.38$\pm$0.43} & {13800$\pm$1740} & {7.90$\pm$0.14} & 	{0.55$\pm$0.08} \\
{2410524453277760640} & {J2309$-$1337} & {23:09:15.58} & {$-$13:37:31.80} & {19.91} & {2.61$\pm$0.62} & {12310$\pm$460} & {8.77$\pm$0.07} & 	{1.09$\pm$0.05} \\
{2379902710847010688} & {J2312$-$2605} & {23:12:03.50} & {$-$26:05:57.41} & {19.22} & {3.64$\pm$0.42} & {13380$\pm$820} & {8.58$\pm$0.10} & 	{0.98$\pm$0.07} \\
{2409194525244964608} & {J2325$-$1312} & {23:25:18.83} & {$-$13:12:44.50} & {19.72} & {2.01$\pm$0.54} & {13900$\pm$1910} & {7.86$\pm$0.17} & 	{0.53$\pm$0.09} \\
{2381034417549322240} & {J2330$-$2449} & {23:30:50.62} & {$-$24:49:52.10} & {20.01} & {3.83$\pm$0.84} & {15070$\pm$820} & {7.69$\pm$0.15} & 	{0.46$\pm$0.07} \\
{2433615060749989760} & {J2331$-$1115} & {23:31:10.02} & {$-$11:15:42.43} & {17.40} & {3.72$\pm$0.17} & {22090$\pm$1030} & {7.99$\pm$0.15} & 	{0.62$\pm$0.08} \\
{2331020725978650624} & {J2334$-$2819} & {23:34:07.41} & {$-$28:19:19.24} & {18.24} & \multicolumn{1}{c}{\nodata} & {12330$\pm$470} & {8.10$\pm$0.11} & 	{0.67$\pm$0.07} \\
{2332573098957941376} & {J2339$-$2531} & {23:39:48.06} & {$-$25:31:18.08} & {17.06} & {$-$0.16$\pm$0.20} & {11750$\pm$430} & {8.13$\pm$0.12} & 	{0.68$\pm$0.08} \\
{2387052560084561024} & {J2340$-$2345} & {23:40:46.46} & {$-$23:45:53.42} & {19.86} & {1.85$\pm$0.84} & {10310$\pm$340} & {8.02$\pm$0.23} & 	{0.61$\pm$0.14} \\
{2339360418595723264} & {J2345$-$2334} & {23:45:39.19} & {$-$23:34:50.63} & {19.07} & {4.03$\pm$0.48} & {12610$\pm$780} & {7.94$\pm$0.19} & 	{0.57$\pm$0.11} \\
{2328335718583448064} & {J2349$-$2743} & {23:49:16.24} & {$-$27:43:26.18} & {17.62} & {0.10$\pm$0.13} & {14590$\pm$1120} & {8.47$\pm$0.10} & 	{0.91$\pm$0.07} \\
\enddata
\vspace{-0.8cm}
\end{deluxetable*}
\label{table:atlas_wd_table}

\startlongtable
\begin{deluxetable*}{llrrcrrrr}
\tablecolumns{9}
\tablewidth{0pt}
\tablecaption{Table of 29 sdA + ELM candidates observed as a part of our ATLAS+SkyMapper target selection.}
\tablehead{
\colhead{Gaia Source ID} & \colhead{Object} & \colhead{R.A.} & \colhead{Dec.} & \colhead{Gaia G} & {Gaia Parallax} & \colhead{T$_{\text{eff}}$} &  \colhead{log(g)} \\
\colhead{} & \colhead{} & \colhead{} & \colhead{} & \colhead{(mag)} & \colhead{(mas)} & \colhead{(K)} & \colhead{(cm s$^{-2}$)} }
\startdata
{4684237675440128512} & {J0019$-$7620} & {00:19:45.64} & {$-$76:20:50.00} & {16.67} & { 0.05$\pm$0.06} & {7520$\pm$70} & {5.02$\pm$0.20} \\
{5006336308261344000} & {J0049$-$3354} & {00:49:30.74} & {$-$33:54:04.68} & {16.57} & { 0.13$\pm$0.08} & {8860$\pm$50} & {5.53$\pm$0.09} \\
{2470088090531076096} & {J0107$-$1042} & {01:07:12.71} & {$-$10:42:35.91} & {20.14} & { 1.38$\pm$0.66} & {9460$\pm$43} & {6.06$\pm$1.34} \\
{4957677283735582336} & {J0155$-$4148} & {01:55:34.86} & {$-$41:48:18.44} & {15.75} & { 2.08$\pm$0.04} & {10060$\pm$70} & {5.61$\pm$0.06} \\
{4942238319415762048} & {J0155$-$4708} & {01:55:53.66} & {$-$47:08:22.61} & {16.93} & { 0.07$\pm$0.07} & {9040$\pm$40} & {5.41$\pm$0.08} \\
{5063798847514778880} & {J0239$-$3157} & {02:39:20.40} & {$-$31:57:06.26} & {15.70} & { 0.14$\pm$0.06} & {8850$\pm$50} & {5.57$\pm$0.08} \\
{5048200350928561792} & {J0313$-$3406} & {03:13:10.01} & {$-$34:06:18.11} & {17.47} & { 0.04$\pm$0.10} & {7940$\pm$50} & {5.06$\pm$0.13} \\
{4838360480913152768} & {J0402$-$4452} & {04:02:01.05} & {$-$44:52:56.01} & {14.57} & { 0.27$\pm$0.04} & {9010$\pm$20} & {5.51$\pm$0.03} \\
{3200233905240195968} & {J0441$-$0547} & {04:41:32.62} & {$-$05:47:34.93} & {18.83} & { 0.31$\pm$0.29} & {8640$\pm$80} & {5.27$\pm$0.17} \\
{3777278773096451712} & {J1046$-$0425} & {10:46:02.98} & {$-$04:25:17.51} & {20.17} & {$-$1.27$\pm$2.07} & {8470$\pm$90} & {5.88$\pm$0.18} \\
{3801357051247738112} & {J1051$-$0347} & {10:51:27.38} & {$-$03:47:09.20} & {18.95} & {$-$0.29$\pm$0.32} & {8960$\pm$80} & {5.49$\pm$0.13} \\
{3599721286026267264} & {J1144$-$0450} & {11:44:20.31} & {$-$04:50:10.39} & {20.01} & { 1.28$\pm$0.73} & {9290$\pm$110} & {5.83$\pm$0.29} \\
{3595644434349398272} & {J1159$-$0633} & {11:59:35.70} & {$-$06:33:46.98} & {19.38} & { 0.29$\pm$0.59} & {11960$\pm$510} & {5.41$\pm$0.16} \\
{3628140710962656512} & {J1308$-$0733} & {13:08:57.86} & {$-$07:33:56.63} & {18.35} & { 0.33$\pm$0.18} & {8590$\pm$80} & {6.15$\pm$0.16} \\
{3643468379794415104} & {J1404$-$0634} & {14:04:40.81} & {$-$06:34:26.62} & {19.79} & { 1.06$\pm$0.92} & {8880$\pm$70} & {5.52$\pm$0.11} \\
{3644329881514712576} & {J1405$-$0455} & {14:05:44.40} & {$-$04:55:19.85} & {19.65} & {$-$0.50$\pm$0.79} & {8990$\pm$120} & {5.53$\pm$0.32} \\
{3641707894175473024} & {J1425$-$0508} & {14:25:55.01} & {$-$05:08:08.60} & {16.39} & { 0.25$\pm$0.08} & {8570$\pm$10} & {5.59$\pm$0.02} \\
{6281296211912043904} & {J1455$-$1858} & {14:55:32.94} & {$-$18:58:01.40} & {17.58} & {$-$0.01$\pm$0.25} & {8760$\pm$90} & {5.93$\pm$0.14} \\
{6305569103622390528} & {J1500$-$1727} & {15:00:01.98} & {$-$17:27:39.13} & {16.29} & { 0.18$\pm$0.10} & {8560$\pm$60} & {5.79$\pm$0.10} \\
{6332713533156000384} & {J1503$-$0750} & {15:03:22.21} & {$-$07:50:24.68} & {20.33} & {$-$0.30$\pm$0.88} & {8590$\pm$110} & {5.48$\pm$0.21} \\
{6334668842786315648} & {J1507$-$0606} & {15:07:17.86} & {$-$06:06:18.22} & {20.32} & { 0.51$\pm$0.82} & {7720$\pm$120} & {5.52$\pm$0.25} \\
{6258208494955477888} & {J1522$-$1737} & {15:22:20.54} & {$-$17:37:50.56} & {20.41} & {$-$0.17$\pm$0.86} & {9090$\pm$140} & {6.01$\pm$0.29} \\
{6844758095371132800} & {J2135$-$1137} & {21:35:47.07} & {$-$11:37:54.19} & {15.76} & { 0.27$\pm$0.07} & {8670$\pm$50} & {5.63$\pm$0.08} \\
{6592840907497444608} & {J2145$-$3135} & {21:45:01.84} & {$-$31:35:57.05} & {18.53} & { 0.31$\pm$0.22} & {7580$\pm$80} & {5.11$\pm$0.20} \\
{6810469138063674624} & {J2151$-$2645} & {21:51:41.01} & {$-$26:45:03.13} & {18.71} & {$-$0.04$\pm$0.25} & {8670$\pm$50} & {5.91$\pm$0.07} \\
{6587685125675331840} & {J2153$-$3630} & {21:53:01.62} & {$-$36:30:03.92} & {17.12} & { 0.27$\pm$0.19} & {8760$\pm$50} & {5.35$\pm$0.11} \\
{6628443918638517632} & {J2226$-$2137} & {22:26:45.58} & {$-$21:37:50.23} & {17.76} & {$-$0.03$\pm$0.15} & {8940$\pm$60} & {5.18$\pm$0.14} \\
{6624542164187962368} & {J2238$-$2333} & {22:38:49.31} & {$-$23:33:08.32} & {18.54} & { 0.32$\pm$0.32} & {11440$\pm$820} & {6.67$\pm$0.33} \\
{2334416331419048576} & {J2354$-$2706} & {23:54:29.97} & {$-$27:06:26.14} & {20.12} & {$-$1.67$\pm$0.93} & {12220$\pm$2330} & {6.99$\pm$0.30} \\
\enddata
\vspace{-0.8cm}
\end{deluxetable*}

\startlongtable
\begin{deluxetable*}{llrrcrrrr}
\tablecolumns{9}
\tablewidth{0pt}
\tablecaption{Table of 49 White Dwarfs identified through our Gaia parallax target selection.
Targets marked with a $^\star$ are also included in Pelisoli \& Vos (2019) as ELM candidates.
All targets listed here are present in the Gaia DR2 white dwarf catalogue of Gentile Fusillo et al. (2019).}
\tablehead{
\colhead{Gaia Source ID} & \colhead{Object} & \colhead{R.A.} & \colhead{Dec.} & \colhead{Gaia G} & \colhead{Gaia Parallax} & \colhead{T$_{\text{eff}}$} &  \colhead{log(g)} & \colhead{M$_{WD}$} \\
\colhead{} & \colhead{} & \colhead{} & \colhead{} & \colhead{(mag)} & \colhead{(mas)} & \colhead{(K)} & \colhead{(cm s$^{-2}$)} & \colhead{(M$_\odot$)} }
\startdata
{4876689387538123008} & {J0455$-$2928 $^\star$} & {04:55:35.93} & {$-$29:28:58.74} & {15.03} & {10.26$\pm$0.03} & {26130$\pm$240} & {7.73$\pm$0.04} & {0.50$\pm$0.03} \\
{4800596031773794944} & {J0518$-$4336} & {05:18:26.98} & {$-$43:36:18.40} & {18.09} & {1.04$\pm$0.15} & {42120$\pm$4070} & {8.61$\pm$0.08} & {1.02$\pm$0.05} \\
{2967020552620016768} & {J0545$-$1902} & {05:45:45.30} & {$-$19:02:45.50} & {17.34} & {2.46$\pm$0.09} & {22610$\pm$200} & {7.42$\pm$0.03} & {0.41$\pm$0.02} \\
{5482174218861274624} & {J0611$-$6044} & {06:11:51.46} & {$-$60:44:22.86} & {17.69} & {2.99$\pm$0.21} & {19690$\pm$230} & {8.08$\pm$0.04} & {0.67$\pm$0.03} \\
{5550454165824297856} & {J0619$-$4942} & {06:19:04.95} & {$-$49:42:37.20} & {18.49} & {1.06$\pm$0.15} & {41850$\pm$3170} & {8.15$\pm$0.07} & {0.76$\pm$0.05} \\
{5555707774117176192} & {J0631$-$4541} & {06:31:44.74} & {$-$45:41:22.20} & {18.07} & {2.09$\pm$0.14} & {34780$\pm$280} & {8.27$\pm$0.06} & {0.81$\pm$0.04} \\
{5484929251404350592} & {J0652$-$5630 $^\star$} & {06:52:56.74} & {$-$56:30:46.40} & {18.45} & {1.70$\pm$0.15} & {27960$\pm$270} & {8.05$\pm$0.05} & {0.67$\pm$0.04} \\
{5483936186245586432} & {J0700$-$5711} & {07:00:56.49} & {$-$57:11:01.60} & {18.58} & {1.52$\pm$0.17} & {24390$\pm$370} & {7.67$\pm$0.06} & {0.47$\pm$0.03} \\
{5281105393618729600} & {J0711$-$6727} & {07:11:01.22} & {$-$67:27:25.20} & {18.23} & {1.52$\pm$0.13} & {38020$\pm$460} & {8.15$\pm$0.07} & {0.75$\pm$0.05} \\
{5288476833106664064} & {J0727$-$6352} & {07:27:28.80} & {$-$63:52:17.20} & {18.31} & {1.63$\pm$0.12} & {37750$\pm$460} & {8.16$\pm$0.07} & {0.76$\pm$0.05} \\
{5210251837830078208} & {J0831$-$7717 $^\star$} & {08:31:14.84} & {$-$77:17:37.40} & {18.07} & {1.26$\pm$0.11} & {41720$\pm$3700} & {8.18$\pm$0.09} & {0.78$\pm$0.06} \\
{5761818336913460096} & {J0847$-$0424} & {08:47:47.66} & {$-$04:24:48.50} & {18.38} & {1.81$\pm$0.19} & {23130$\pm$260} & {7.63$\pm$0.04} & {0.47$\pm$0.02} \\
{5196645102262353152} & {J0902$-$8034 $^\star$} & {09:02:52.09} & {$-$80:34:52.80} & {18.56} & {1.33$\pm$0.15} & {32410$\pm$440} & {7.64$\pm$0.11} & {0.49$\pm$0.04} \\
{5660008787157868928} & {J0950$-$2511 $^\star$} & {09:50:17.95} & {$-$25:11:25.30} & {15.53} & {3.16$\pm$0.06} & {36580$\pm$230} & {7.47$\pm$0.04} & {0.44$\pm$0.02} \\
{5688043614950782848} & {J0955$-$1209 $^\star$} & {09:55:58.27} & {$-$12:09:37.30} & {17.89} & {2.16$\pm$0.22} & {26710$\pm$270} & {8.01$\pm$0.04} & {0.64$\pm$0.03} \\
{3752200596493839232} & {J1024$-$1434 $^\star$} & {10:24:32.21} & {$-$14:34:20.50} & {16.84} & {2.54$\pm$0.17} & {22240$\pm$300} & {7.21$\pm$0.04} & {0.36$\pm$0.01} \\
{3762701108633074688} & {J1040$-$0746 $^\star$} & {10:40:26.09} & {$-$07:46:14.97} & {17.76} & {2.16$\pm$0.16} & {22710$\pm$200} & {7.53$\pm$0.03} & {0.43$\pm$0.02} \\
{3565938482025538944} & {J1111$-$1213} & {11:11:14.66} & {$-$12:13:11.20} & {18.43} & {1.01$\pm$0.26} & {28750$\pm$300} & {7.23$\pm$0.05} & {0.39$\pm$0.01} \\
{5397661047866038400} & {J1124$-$3752} & {11:24:55.89} & {$-$37:52:28.30} & {18.12} & {1.92$\pm$0.18} & {34390$\pm$300} & {8.10$\pm$0.06} & {0.71$\pm$0.04} \\
{3480932145705503616} & {J1149$-$2852 $^\star$} & {11:49:51.97} & {$-$28:52:39.70} & {18.17} & {1.99$\pm$0.18} & {16670$\pm$230} & {7.53$\pm$0.04} & {0.42$\pm$0.02} \\
{3695067154816052736} & {J1222$-$0106} & {12:22:06.38} & {$-$01:06:36.20} & {18.37} & {1.84$\pm$0.23} & {28160$\pm$510} & {8.12$\pm$0.09} & {0.71$\pm$0.06} \\
{3467973099797096192} & {J1226$-$3408} & {12:26:00.06} & {$-$34:08:25.10} & {18.49} & {1.12$\pm$0.32} & {22590$\pm$350} & {7.33$\pm$0.05} & {0.40$\pm$0.02} \\
{6156485046406365568} & {J1235$-$3745} & {12:35:44.76} & {$-$37:45:00.00} & {17.27} & {2.87$\pm$0.15} & {28010$\pm$470} & {7.58$\pm$0.08} & {0.47$\pm$0.03} \\
{3707477754875428352} & {J1239$+$0514} & {12:39:32.01} & {$+$05:14:07.80} & {18.43} & {1.47$\pm$0.22} & {45580$\pm$1280} & {7.94$\pm$0.18} & {0.65$\pm$0.09} \\
{3504361849435922048} & {J1307$-$2207 $^\star$} & {13:07:42.65} & {$-$22:07:40.70} & {17.30} & {2.06$\pm$0.12} & {34120$\pm$260} & {8.00$\pm$0.05} & {0.66$\pm$0.04} \\
{3624296749592496768} & {J1312$-$1025} & {13:12:39.30} & {$-$10:25:53.30} & {18.01} & {1.78$\pm$0.19} & {24200$\pm$250} & {7.85$\pm$0.04} & {0.55$\pm$0.03} \\
{6183015574790579328} & {J1319$-$2844 $^\star$} & {13:19:07.72} & {$-$28:44:07.70} & {18.36} & {1.35$\pm$0.19} & {33330$\pm$290} & {7.82$\pm$0.06} & {0.57$\pm$0.03} \\
{3636151129911425408} & {J1319$-$0413} & {13:19:09.85} & {$-$04:13:14.13} & {17.44} & {2.40$\pm$0.32} & {11330$\pm$120} & {8.02$\pm$0.04} & {0.61$\pm$0.03} \\
{3607395533590721024} & {J1319$-$1639} & {13:19:42.20} & {$-$16:39:09.20} & {18.58} & {1.57$\pm$0.24} & {28320$\pm$400} & {8.32$\pm$0.07} & {0.57$\pm$0.03} \\
{3610368033211697024} & {J1338$-$1211} & {13:38:59.68} & {$-$12:11:04.20} & {17.68} & {1.92$\pm$0.16} & {41710$\pm$2810} & {8.21$\pm$0.07} & {0.79$\pm$0.05} \\
{6115720618448356736} & {J1349$-$3652} & {13:49:42.74} & {$-$36:52:31.90} & {17.75} & {1.97$\pm$0.20} & {25480$\pm$390} & {7.68$\pm$0.06} & {0.48$\pm$0.03} \\
{6290699097913837696} & {J1355$-$1946} & {13:55:41.73} & {$-$19:46:31.20} & {17.74} & {2.45$\pm$0.23} & {31940$\pm$260} & {8.13$\pm$0.06} & {0.73$\pm$0.04} \\
{6120865237649732352} & {J1406$-$3622} & {14:06:01.84} & {$-$36:22:29.70} & {16.71} & {3.77$\pm$0.08} & {21900$\pm$410} & {8.07$\pm$0.06} & {0.67$\pm$0.04} \\
{6122949361939805184} & {J1411$-$3434} & {14:11:28.59} & {$-$34:34:41.20} & {17.28} & {2.03$\pm$0.13} & {42430$\pm$8080} & {8.38$\pm$0.16} & {0.89$\pm$0.10} \\
{6283002069842683392} & {J1426$-$2006} & {14:26:25.06} & {$-$20:06:30.70} & {18.54} & {0.84$\pm$0.48} & {42150$\pm$7540} & {9.13$\pm$0.14} & {1.27$\pm$0.07} \\
{6324871678787667968} & {J1436$-$1106} & {14:36:13.35} & {$-$11:06:48.19} & {17.79} & {1.58$\pm$0.26} & {22150$\pm$290} & {7.44$\pm$0.04} & {0.41$\pm$0.02} \\
{6284835750296054784} & {J1436$-$1845} & {14:38:58.74} & {$-$18:45:42.70} & {16.59} & {3.08$\pm$0.12} & {42400$\pm$5400} & {8.88$\pm$0.08} & {1.16$\pm$0.04} \\
{6338239094480660608} & {J1442$-$0352} & {14:42:55.33} & {$-$03:52:10.70} & {17.27} & {2.92$\pm$0.13} & {35250$\pm$330} & {8.20$\pm$0.07} & {0.77$\pm$0.05} \\
{6227876645636436352} & {J1503$-$2348 $^\star$} & {15:03:55.16} & {$-$23:48:27.70} & {18.12} & {1.22$\pm$0.19} & {34550$\pm$430} & {7.96$\pm$0.08} & {0.64$\pm$0.05} \\
{6318961082233123456} & {J1507$-$1028 $^\star$} & {15:07:28.33} & {$-$10:28:59.10} & {18.21} & {1.03$\pm$0.20} & {31300$\pm$230} & {7.55$\pm$0.05} & {0.46$\pm$0.02} \\
{6253056939379854976} & {J1518$-$2047} & {15:18:04.69} & {$-$20:47:01.80} & {18.57} & {1.35$\pm$0.78} & {41780$\pm$7810} & {6.77$\pm$0.22} & {0.28$\pm$0.05} \\
{6243072519105238272} & {J1611$-$2117} & {16:11:31.51} & {$-$21:17:37.60} & {17.27} & {2.13$\pm$0.14} & {32050$\pm$380} & {7.29$\pm$0.08} & {0.40$\pm$0.03} \\
{4334088989163587200} & {J1645$-$1127} & {16:45:37.40} & {$-$11:27:54.60} & {18.09} & {1.96$\pm$0.18} & {42190$\pm$7910} & {8.76$\pm$0.13} & {1.10$\pm$0.07} \\
{5766955358317857280} & {J1650$-$8614} & {16:50:45.73} & {$-$86:14:35.90} & {16.52} & {2.93$\pm$0.07} & {31680$\pm$260} & {7.52$\pm$0.05} & {0.46$\pm$0.02} \\
{4385335984889341696} & {J1655$+$0306 $^\star$} & {16:55:54.57} & {$+$03:06:11.40} & {16.60} & {2.54$\pm$0.08} & {44410$\pm$1130} & {7.57$\pm$0.13} & {0.50$\pm$0.05} \\
{4381363208860631296} & {J1700$+$0044 $^\star$} & {17:00:08.87} & {$+$00:44:36.30} & {18.45} & {1.26$\pm$0.21} & {31270$\pm$340} & {8.10$\pm$0.07} & {0.61$\pm$0.05} \\
{4392410380142049792} & {J1700$+$0512 $^\star$} & {17:00:56.61} & {$+$05:12:54.60} & {17.59} & {1.81$\pm$0.12} & {42080$\pm$4150} & {8.30$\pm$0.08} & {0.85$\pm$0.05} \\
{4390681746002706176} & {J1726$+$0601} & {17:26:31.37} & {$+$06:01:00.90} & {16.09} & {3.63$\pm$0.10} & {41960$\pm$2870} & {8.29$\pm$0.06} & {0.84$\pm$0.04} \\
{4389028630270429440} & {J1728$+$0354} & {17:28:18.00} & {$+$03:54:55.30} & {16.61} & {3.27$\pm$0.09} & {28500$\pm$430} & {7.46$\pm$0.07} & {0.43$\pm$0.03} \\
\enddata
\vspace{-0.8cm}
\end{deluxetable*}

\startlongtable
\begin{deluxetable*}{llrrcrrrr}
\tablecolumns{9}
\tablewidth{0pt}
\tablecaption{Table of seven sdB stars observed as a part of our Gaia parallax target selection.
Atmosphere parameters are based on pure-Hydrogen atmosphere model fits, and should be used with caution.
Targets marked with a $^\dagger$ are present in the Geier et al. (2019) Gaia hot subluminous star catalogue.
Targets marked with a $^\star$ are present in Pelisoli et al. (2019) as ELM candidates based on Gaia colors.}
\tablehead{
\colhead{Gaia Source ID} & \colhead{Object} & \colhead{R.A.} & \colhead{Dec.} & \colhead{Gaia G} & {Gaia Parallax} & \colhead{T$_{\text{eff}}$} &  \colhead{log(g)} \\
\colhead{} & \colhead{} & \colhead{} & \colhead{} & \colhead{(mag)} & \colhead{(mas)} & \colhead{(K)} & \colhead{(cm s$^{-2}$)} }
\startdata
{5453140446099189120} & {J1054$-$2941} & {10:54:53.64} & {$-$29:41:10.24} & {17.01} & {0.48$\pm$0.10} & {34020$\pm$420} & {5.57$\pm$0.09} \\
{3548810053666523904} & {J1137$-$1447 $^\dagger$$^\star$} & {11:37:26.73} & {$-$14:47:57.10} & {16.35} & {0.86$\pm$0.11} & {29430$\pm$230} & {5.64$\pm$0.04} \\
{3470421329940244608} & {J1231$-$3104 $^\dagger$} & {12:31:29.71} & {$-$31:04:31.20} & {18.45} & {1.15$\pm$0.28} & {31120$\pm$520} & {5.79$\pm$0.11} \\
{6173348947732101504} & {J1359$-$3054} & {13:59:17.78} & {$-$30:54:09.61} & {16.63} & {0.94$\pm$0.22} & {30810$\pm$320} & {5.55$\pm$0.06} \\
{6322166948902534016} & {J1517$-$0706 $^\dagger$} & {15:17:59.87} & {$-$07:06:02.80} & {18.09} & {0.91$\pm$0.27} & {36580$\pm$500} & {6.06$\pm$0.09} \\
{6322703991612652160} & {J1523$-$0609 $^\dagger$} & {15:23:51.71} & {$-$06:09:35.40} & {18.59} & {1.55$\pm$0.45} & {28850$\pm$380} & {5.44$\pm$0.06} \\
{4353523544382401408} & {J1648$-$0447 $^\dagger$$^\star$} & {16:48:06.27} & {$-$04:47:25.30} & {15.51} & {1.30$\pm$0.06} & {33180$\pm$170} & {6.20$\pm$0.04} \\
\enddata
\vspace{1cm}
\end{deluxetable*}
\label{table:gaia_sdb_table}

\startlongtable
\begin{deluxetable*}{crr}
\tablecolumns{3}
\tablewidth{0pt}
\tablecaption{Radial velocity data.}
\tablehead{
\colhead{Object} & \colhead{HJD} & \colhead{v$_{helio}$} \\
\colhead{} & \colhead{($-$2450000 d)} & \colhead{(km s$^{-1}$)} }
\startdata
{J0500$-$0930} & {8401.926694} & {$-186.75\pm13.65$} \\
{\nodata} & {8457.747110} & {$61.62\pm12.35$} \\
{\nodata} & {8457.750212} & {$87.93\pm17.82$} \\
{\nodata} & {8457.752365} & {$64.62\pm11.47$} \\
{\nodata} & {8457.754518} & {$74.18\pm7.99$} \\
{\nodata} & {8457.756659} & {$71.84\pm12.02$} \\
{\nodata} & {8457.759159} & {$53.42\pm9.72$} \\
{\nodata} & {8457.763418} & {$61.53\pm13.60$} \\
{\nodata} & {8457.765744} & {$42.45\pm10.87$} \\
{\nodata} & {8457.768071} & {$36.62\pm11.64$} \\
{\nodata} & {8457.770224} & {$26.41\pm8.28$} \\
{\nodata} & {8457.772376} & {$38.68\pm11.72$} \\
{\nodata} & {8457.775386} & {$29.18\pm15.14$} \\
{\nodata} & {8457.777701} & {$25.29\pm12.01$} \\
{\nodata} & {8457.779853} & {$27.02\pm16.79$} \\
{\nodata} & {8457.782006} & {$44.66\pm21.64$} \\
{\nodata} & {8457.784159} & {$28.22\pm16.39$} \\
{\nodata} & {8457.787226} & {$9.46\pm11.72$} \\
{\nodata} & {8457.789552} & {$-20.99\pm17.46$} \\
{\nodata} & {8457.792052} & {$-5.07\pm15.14$} \\
{\nodata} & {8457.795073} & {$-12.70\pm13.12$} \\
{\nodata} & {8457.798094} & {$-18.85\pm15.36$} \\
{\nodata} & {8457.803210} & {$-33.28\pm8.90$} \\
{\nodata} & {8457.808140} & {$-26.66\pm14.57$} \\
{\nodata} & {8457.813418} & {$-42.91\pm11.08$} \\
{\nodata} & {8457.818349} & {$-78.66\pm10.61$} \\
{\nodata} & {8457.822585} & {$-76.56\pm8.71$} \\
{\nodata} & {8457.827237} & {$-95.39\pm8.77$} \\
{\nodata} & {8457.830779} & {$-87.40\pm11.60$} \\
{\nodata} & {8457.834321} & {$-100.28\pm13.18$} \\
{\nodata} & {8457.837851} & {$-103.77\pm9.88$} \\
{\nodata} & {8457.841393} & {$-112.61\pm14.57$} \\
{\nodata} & {8457.845941} & {$-129.67\pm11.71$} \\
{\nodata} & {8457.849483} & {$-125.85\pm10.11$} \\
{\nodata} & {8457.853025} & {$-132.19\pm12.94$} \\
{\nodata} & {8457.856566} & {$-157.25\pm10.30$} \\
{\nodata} & {8457.860096} & {$-170.16\pm17.74$} \\
{\nodata} & {8457.871103} & {$-170.80\pm8.32$} \\
{\nodata} & {8457.877388} & {$-179.44\pm7.53$} \\
{\nodata} & {8457.881624} & {$-183.02\pm12.54$} \\
{\nodata} & {8457.885675} & {$-191.17\pm10.47$} \\
{\nodata} & {8457.891717} & {$-175.41\pm13.15$} \\
{\nodata} & {8463.732241} & {$-29.33\pm29.10$} \\
{\nodata} & {8463.743653} & {$-113.25\pm17.41$} \\
{\nodata} & {8463.754995} & {$-118.16\pm13.77$} \\
{\nodata} & {8463.766153} & {$-139.52\pm14.72$} \\
{\nodata} & {8463.846244} & {$-153.90\pm28.09$} \\
{\nodata} & {8463.880908} & {$-143.56\pm14.28$} \\
{\nodata} & {8483.649329} & {$-5.27\pm7.39$} \\
{\nodata} & {8487.778842} & {$-70.86\pm16.18$} \\
{\nodata} & {8487.877345} & {$-168.87\pm16.41$} \\
{\nodata} & {8543.539722} & {$-130.75\pm5.38$} \\
{\nodata} & {8544.502663} & {$92.41\pm6.48$} \\
{\nodata} & {8544.570195} & {$-43.28\pm6.47$} \\
{\nodata} & {8545.505076} & {$-91.40\pm11.20$} \\
{\nodata} & {8545.587750} & {$46.63\pm7.20$} \\
{\nodata} & {8546.509416} & {$29.89\pm5.14$} \\
{\nodata} & {8546.556205} & {$-100.47\pm5.65$} \\
\hline
{J0642$-$5605} & {8543.572777} & {$5.44\pm8.51$} \\
{\nodata} & {8544.533324} & {$602.87\pm5.99$} \\
{\nodata} & {8544.582158} & {$125.65\pm6.20$} \\
{\nodata} & {8545.542634} & {$-88.83\pm7.63$} \\
{\nodata} & {8545.569443} & {$310.89\pm6.79$} \\
{\nodata} & {8545.599940} & {$585.30\pm7.48$} \\
{\nodata} & {8545.644891} & {$18.54\pm5.38$} \\
{\nodata} & {8545.687739} & {$102.11\pm18.99$} \\
{\nodata} & {8546.521206} & {$599.44\pm6.90$} \\
{\nodata} & {8546.596143} & {$-104.66\pm7.31$} \\
{\nodata} & {8546.597971} & {$-82.35\pm7.09$} \\
{\nodata} & {8546.599798} & {$-62.04\pm9.20$} \\
{\nodata} & {8546.615714} & {$159.24\pm8.57$} \\
{\nodata} & {8546.665951} & {$517.46\pm8.47$} \\
\hline
{J0650$-$4926} & {8543.587096} & {$-134.81\pm8.78$} \\
{\nodata} & {8544.540141} & {$355.19\pm8.86$} \\
{\nodata} & {8544.587890} & {$11.27\pm6.11$} \\
{\nodata} & {8545.549203} & {$211.29\pm11.40$} \\
{\nodata} & {8545.576119} & {$405.68\pm10.41$} \\
{\nodata} & {8545.605456} & {$288.43\pm13.11$} \\
{\nodata} & {8545.649962} & {$-148.42\pm9.08$} \\
{\nodata} & {8545.711134} & {$63.90\pm26.51$} \\
{\nodata} & {8546.527364} & {$-138.46\pm10.98$} \\
{\nodata} & {8546.587008} & {$114.24\pm11.71$} \\
{\nodata} & {8546.589183} & {$139.26\pm10.67$} \\
{\nodata} & {8546.591357} & {$207.90\pm9.54$} \\
{\nodata} & {8546.621214} & {$386.67\pm13.79$} \\
\hline
{J0930$-$8107} & {8543.662882} & {$-160.79\pm12.12$} \\
{\nodata} & {8544.545298} & {$-155.73\pm15.86$} \\
{\nodata} & {8544.592581} & {$147.79\pm12.36$} \\
{\nodata} & {8545.556906} & {$65.29\pm9.17$} \\
{\nodata} & {8545.581480} & {$166.37\pm12.71$} \\
{\nodata} & {8545.609496} & {$-190.91\pm12.01$} \\
{\nodata} & {8545.654017} & {$192.66\pm12.16$} \\
{\nodata} & {8546.531874} & {$116.24\pm17.23$} \\
{\nodata} & {8546.575733} & {$-134.02\pm12.31$} \\
{\nodata} & {8546.577908} & {$-112.28\pm12.64$} \\
{\nodata} & {8546.580083} & {$-161.69\pm13.61$} \\
{\nodata} & {8546.604104} & {$-110.32\pm11.41$} \\
{\nodata} & {8546.625537} & {$196.95\pm16.17$} \\
{\nodata} & {8546.804837} & {$194.15\pm14.42$} \\
\hline
{J1236$-$0444} & {7835.658070} & {$194.99\pm20.79$} \\
{\nodata} & {7835.682442} & {$199.69\pm27.99$} \\
{\nodata} & {7835.740827} & {$128.05\pm26.07$} \\
{\nodata} & {7835.800865} & {$-1.00\pm24.97$} \\
{\nodata} & {7835.848070} & {$-11.05\pm29.47$} \\
{\nodata} & {7836.703611} & {$5.86\pm22.03$} \\
{\nodata} & {7836.806389} & {$106.27\pm24.03$} \\
{\nodata} & {7836.809267} & {$97.28\pm23.08$} \\
{\nodata} & {7836.858969} & {$140.20\pm25.13$} \\
{\nodata} & {8187.657646} & {$225.69\pm28.23$} \\
{\nodata} & {8187.736842} & {$178.40\pm25.95$} \\
{\nodata} & {8189.750132} & {$226.14\pm19.57$} \\
{\nodata} & {8189.823680} & {$142.52\pm19.38$} \\
{\nodata} & {8190.686871} & {$-38.93\pm26.37$} \\
{\nodata} & {8190.751398} & {$-12.03\pm25.62$} \\
{\nodata} & {8191.652700} & {$184.28\pm27.53$} \\
{\nodata} & {8191.722922} & {$240.31\pm26.46$} \\
\hline
{J1425$-$0508} & {7835.716718} & {$-101.51\pm13.89$} \\
{\nodata} & {7835.772831} & {$-35.31\pm10.99$} \\
{\nodata} & {7835.832876} & {$-82.13\pm12.91$} \\
{\nodata} & {7835.896018} & {$-27.08\pm11.05$} \\
{\nodata} & {7835.899585} & {$-23.59\pm10.82$} \\
{\nodata} & {7836.738278} & {$-26.88\pm11.12$} \\
{\nodata} & {7836.840432} & {$6.98\pm11.65$} \\
{\nodata} & {7836.883911} & {$19.45\pm11.87$} \\
{\nodata} & {7836.908636} & {$25.57\pm11.91$} \\
{\nodata} & {7929.652855} & {$11.30\pm12.25$} \\
{\nodata} & {7929.656732} & {$6.36\pm12.58$} \\
{\nodata} & {7929.659856} & {$14.39\pm12.28$} \\
{\nodata} & {7929.661766} & {$27.48\pm13.07$} \\
{\nodata} & {8187.770450} & {$-72.60\pm18.95$} \\
{\nodata} & {8187.830644} & {$-62.10\pm19.55$} \\
{\nodata} & {8187.883105} & {$-59.25\pm18.67$} \\
{\nodata} & {8189.673532} & {$38.96\pm16.27$} \\
{\nodata} & {8189.692531} & {$49.73\pm15.86$} \\
{\nodata} & {8189.755714} & {$40.61\pm16.27$} \\
{\nodata} & {8189.828887} & {$79.22\pm16.51$} \\
{\nodata} & {8189.891552} & {$-32.15\pm18.23$} \\
{\nodata} & {8190.719889} & {$-60.65\pm17.61$} \\
{\nodata} & {8190.786916} & {$-50.04\pm17.79$} \\
{\nodata} & {8191.748295} & {$-55.04\pm19.08$} \\
{\nodata} & {8191.831532} & {$-56.06\pm17.58$} \\
\hline
{J1514$-$1436} & {7835.865608} & {$-25.98\pm10.77$} \\
{\nodata} & {7835.906708} & {$-79.80\pm20.57$} \\
{\nodata} & {7835.910275} & {$-47.41\pm16.12$} \\
{\nodata} & {7836.746053} & {$222.44\pm16.32$} \\
{\nodata} & {7836.815246} & {$304.37\pm11.62$} \\
{\nodata} & {7836.845061} & {$283.51\pm17.56$} \\
{\nodata} & {7836.889670} & {$277.96\pm31.33$} \\
{\nodata} & {7931.665442} & {$281.13\pm20.56$} \\
{\nodata} & {8187.837286} & {$171.23\pm21.55$} \\
{\nodata} & {8189.705904} & {$309.81\pm13.32$} \\
{\nodata} & {8189.767503} & {$247.97\pm17.65$} \\
{\nodata} & {8189.841205} & {$187.74\pm14.23$} \\
{\nodata} & {8190.779223} & {$192.88\pm19.40$} \\
{\nodata} & {8190.845175} & {$275.57\pm22.83$} \\
{\nodata} & {8191.761294} & {$-52.28\pm20.39$} \\
{\nodata} & {8191.848146} & {$-61.68\pm23.68$} \\
\enddata
\vspace{-0.8cm}
\end{deluxetable*}
\label{table:rv_table}

\end{document}